\documentclass[useAMS,usenatbib]{mn2e}

\newcommand{\degree}{\ensuremath{^\circ}}

\usepackage{amsmath,natbib,graphicx}
\usepackage{epsf}
\usepackage{epstopdf}
\input{epsf}
\usepackage{epsfig}
\usepackage{color}
\usepackage{subfigure}
\usepackage{caption}

\DeclareGraphicsExtensions{.jpg,.pdf,.png,.eps,.ps}

\newcommand{\HI}{H\:\!\textsc{i}}

\def\simgt{\mathrel{\raise0.35ex\hbox{$\scriptstyle >$}\kern-0.6em
\lower0.40ex\hbox{{$\scriptstyle \sim$}}}}
\def\lta{\mathrel{\raise0.35ex\hbox{$\scriptstyle <$}\kern-0.6em
\lower0.40ex\hbox{{$\scriptstyle \sim$}}}}

\def\gs{\mathrel{\raise0.35ex\hbox{$\scriptstyle >$}\kern-0.6em
\lower0.40ex\hbox{{$\scriptstyle \sim$}}}}
\def\ls{\mathrel{\raise0.35ex\hbox{$\scriptstyle <$}\kern-0.6em
\lower0.40ex\hbox{{$\scriptstyle \sim$}}}}

\title[Molecular gas and metallicity]{Molecular gas as the driver of fundamental galactic relations}
\author[M.\,S.\ Bothwell et al. ]
{M. S. Bothwell$^{1,2}$\thanks{E-mail:
matthew.bothwell@gmail.com},
R. Maiolino$^{1,2}$,
Y. Peng$^{1,2}$,
C. Cicone$^{3}$, 
H. Griffith$^{1,2}$, \newauthor
J. Wagg$^{4}$
\\ 
$^{1}$Cavendish Laboratory, University of Cambridge, 19 J.J. Thomson Avenue, Cambridge, CB3 0HE, UK\\
$^{2}$Kavli Institute for Cosmology, University of Cambridge, Madingley Road, Cambridge CB3 0HA, UK\\
$^{3}$ETH Zurich, Institute for Astronomy, Wolfgang-Pauli-Strasse 27, CH-8093 Zurich, Switzerland\\
$^{4}$SKA Organisation, Lower Withington, UK\\
}
\begin{document}
\date{Accepted ----. Received ---- in original form ----}

\pagerange{\pageref{firstpage}--\pageref{lastpage}} \pubyear{2014}

\maketitle
\begin{abstract}

There has been much recent work dedicated to exploring secondary correlations in the mass-metallicity relation, with significant dependence on both the star formation rate and \HI\  content being demonstrated. 
Previously, a paucity of molecular gas data (combined with sample selection bias) hampered the investigation of any such relation with  molecular gas content. 
In this work, we assemble a sample of 221 galaxies from a variety of surveys in the redshift range $0 < z  < 2$, to explore the connection between molecular gas content and metallicity. 

We explore the effect of gas mass on the mass-metallicity relation, finding that the offset from the relation is negatively correlated against both molecular and total gas mass. We then employ a principle component analysis technique to explore secondary dependences in the mass-metallicity relation, finding that the secondary dependence with gas mass is significantly stronger than with star formation rate, and as such the underlying `Fundamental Metallicity Relation' is between stellar mass, metallicity, and gas mass. In particular, the metallicity dependence on SFR is simply a byproduct of the dependence on the molecular gas content, via the Schmidt-Kennicutt relation. Finally, we note that our principle component analysis finds essentially no connection between gas-phase metallicity and the {\it efficiency} of star formation. 

\end{abstract}


\section{Introduction}
\label{sec:introduction}

Much of the quest to understand the processes driving galaxy evolution has concentrated on understanding the so-called `scaling relations', distinctive correlations existing between large-scale physical parameters of galaxies. These scaling relations allow insight into the mechanisms underlying galaxy formation by providing simple, quantitative tests for models -- both analytical and numerical -- to match. Indeed, the ability (or otherwise) of a model to reproduce a variety of observed scaling relations has become a critical metric by which a model's success is judged.

In recent years, the well-known scaling relations between stellar mass and metallicity (the `mass-metallicity relation'), and stellar mass and star formation rate (the `main sequence' of galaxy evolution) have been extended into a three-dimensional relation, known as the Fundamental Metallicity Relation (FMR; \citealt{2010MNRAS.408.2115M}; \citealt{2010A&A...521L..53L}). Galaxies up to at least $z\sim2.5$ lie on the surface defined by the FMR, on which (a) more massive galaxies have higher metallicities, and (b) at a given stellar mass, galaxies with higher star formation rates have systematically lower metallicities. 

The existence of the FMR has provided an excellent target for models to reproduce, with a variety of models successfully reproducing the general form of the relation, including analytical models (\citealt{2012MNRAS.421...98D}; \citealt{2012arXiv1202.4770D}), semi-analytical models (\citealt{2014MNRAS.439.3817Y}), and SPH/radiative transfer models (\citealt{2014arXiv1404.0043O}). As noted by \cite{2013MNRAS.433.1425B}, these models have primarily explained the correlation between stellar mass, metallicity, and star formation rate in terms of the global behaviour of gas, which can dilute the metallicity of the ISM via pristine inflows, trigger metallicity-enriching bursts of star formation, and remove metals from the ISM in the form of galactic winds.

Based on this, \cite{2013MNRAS.433.1425B} conducted a study of 4253 local galaxies, finding that the mass-metallicity relation also exhibited a significant secondary dependence on mass of atomic hydrogen (\HI), to the extent that the `\HI-FMR' was potentially more fundamental than the correlation with SFR (in terms of the scatter around the respective relations). As with the SFR-FMR before, this has proved a fertile area for models, with several authors concluding that it is indeed the gas content of galaxies that fundamentally underlies and drives the mass-metallicity relation. \cite{2012MNRAS.421...98D} present an analytical model of galaxy evolution, which predicts that metallicity will anti-correlate with {\sc Hi} mass. \citealt{2014ApJ...791..130Z} present a model whereby the metallicity of galaxies is regulated by the gas-to-stellar mass ratio, such that the mass-metallicity relation can be seen as a consequence of the underlying metallicity/gas-to-stellar mass ratio scaling. \citealt{2014arXiv1406.6397A} conduct a similar investigation, extending to resolved studies of local galaxies, and reaching similar conclusions. (See also \citealt{2012MNRAS.427.1075M}; \citealt{2014RMxAC..44R.179L}; \citealt{2015arXiv150102668H}) for additional work).

One potential surprise, however, was the lack of a clear dependence with {\it molecular} gas. While \cite{2013MNRAS.433.1425B} found a close and tight relationship between stellar mass, metallicity, and \HI\ mass, any such relation with H$_2$ mass remained elusive. As discussed by \cite{2013MNRAS.433.1425B}, this was potentially due to two factors: (1) sample selection bias, resulting in a small number of galaxies covering insufficient parameter space, and (2) the fact that measuring H$_2$ mass requires a `conversion factor' (to convert from its observable tracer, $^{12}$CO), which is itself a function of metallicity.  This lack of connection between molecular gas mass and the mass-metallicity relation was particularly surprising, because the physical processes thought to be driving the \HI-FMR (including star formation-triggering inflows, and metal rich outflows) are intimately connected with the molecular gas component of the ISM. Furthermore, it has long been established that molecular gas and star formation are tightly correlated across a wide range of galaxy types, via the Kennicutt-Schmidt relation (\citealt{1959ApJ...129..243S}; \citealt{1998ApJ...498..541K}); given this  relation, a H$_2$-FMR should be expected on the basis of the existence of the SFR-FMR alone. As such, \cite{2013MNRAS.433.1425B} concluded that while a H$_2$-FMR was likely to exist, quantifying its existence was beyond the scope of their data.

In this work, we assemble a large sample of galaxies, from dwarf galaxies in the local Universe to extreme starbursts at $z\sim2$, to explore the connection between the molecular gas content of galaxies, and their gas-phase metallicity. 
First, we examine the effect of molecular gas mass on metallicity scaling relations, by searching for a `H$_2$-FMR' effect, whereby galaxies show offsets from the mass-metallicity relation which correlate with molecular (and total) gas mass. We then adopt a non-parametric statistical approach, and use a principle component analysis technique to examine correlations in stellar mass/metallicity/gas mass parameter space, demonstrating that (a) the most significant `secondary correlation' in the mass-metallicity relation is indeed with gas content, and (b) there is very little correlation between metallicity and star formation efficiency.

Throughout this work we use a \cite{2001MNRAS.322..231K} IMF, and a cosmology following the latest Planck results, with H$_0= (67.8 \pm 0.9)$ km/s/Mpc and $\Omega_m = 0.308 \pm 0.012$ \citep{2015arXiv150201589P}.

\section{Sample selection}
\label{sec:sample}

In this work we will be characterising the scaling relations between a variety of parameters: stellar mass, metallicity, star formation rate, and gas content (both atomic and molecular). We have selected our samples based on the simultaneous availability of all of these parameters. Stellar masses, star formation rates, atomic gas masses, and metallicities (typically measured via optical strong line diagnostics) are comparably easily measured, and are widely available in the literature. Molecular gas masses (measured using observations of $^{12}$CO) are less widely available, and it is the availability of this latter parameter that typically represents the limiting factor in our sample selection. 

In the local Universe, we use three recent surveys for molecular gas which fulfil these criteria excellently: COLD GASS \citep{2011MNRAS.415...32S}, the Herschel Reference Survey \citep{2014arXiv1401.7773B}, and ALLSMOG (Bothwell et al. 2014a). Both COLD GASS and ALLSMOG are selected from the SDSS spectroscopic survey, and as such have available optical spectra for deriving metallicities. Optical spectroscopy for the Herschel Reference Survey (HRS) is available from \cite{2013A&A...550A.114B}. In order to gather better statistics at the low-mass, low-metallicity end of the distribution, we also use galaxies from the volume-limited Local Volume Legacy survey, which (being a complete volume-limited sample within 11 Mpc) contains galaxies of lower mass than appear in other samples. 

We also wish to extend our study beyond the local Universe: in order to compare the samples discussed above to those at higher redshifts, we gather samples of both `normal' star forming galaxies (selected using optical colour cuts, and lying on or around the main sequence) and luminous sub-millimetre galaxies, SMGs (selected via their sum-mm flux, and having extreme star formation rates significantly above the star-forming `main sequence'). These high-$z$ galaxies typically lie around $z\sim2$. 

Below, we briefly describe these various samples. A histogram of stellar masses for galaxies used in this work (separated by sample) is shown in Fig. \ref{fig:hist}. Our combined samples provide a mass coverage of approximately three orders of magnitude, from $8.5 < \log {\rm M*}/{\rm M}_{\sun} < 11.5 $. In total, our combined sample consists of 221 galaxies from $0 < z \lta 2$.

\subsection{Local Universe surveys}

\begin{figure}
\centering
\includegraphics[width=8.5cm]{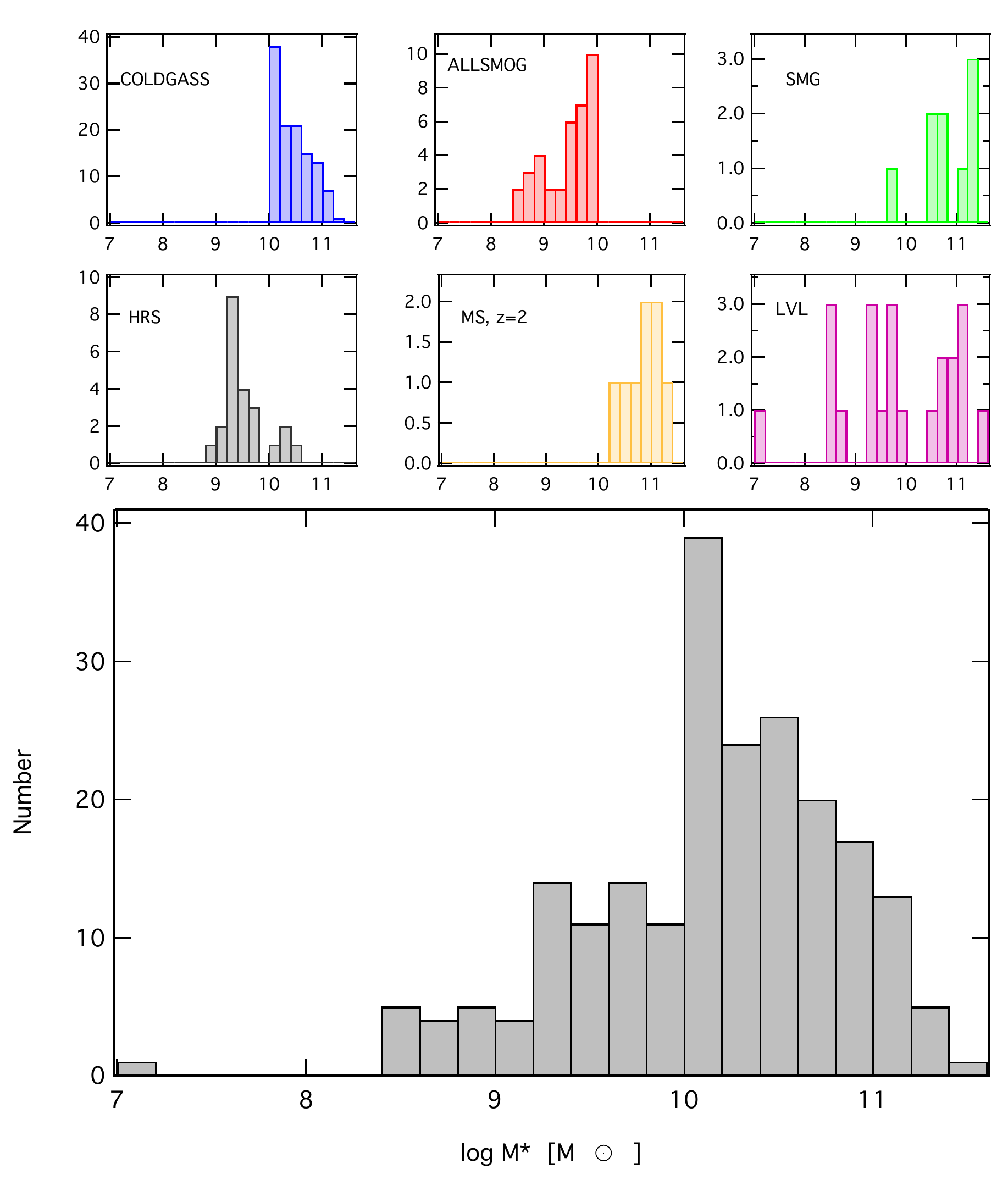}	
\caption{Histogram of the stellar masses of galaxies in the samples used in this work. Only galaxies that obey the selection criteria detailed in \S \ref{sec:sample} are shown. The main panel shows the combined stellar mass histogram, while the smaller inset panels show histograms of the stellar mass for each separate sample.}
\label{fig:hist}
\end{figure}

\subsubsection{ALLSMOG}

ALLSMOG, the Apex Low-redshift Legacy Survey for MOlecular Gas, is an APEX large programme designed to survey molecular gas in local galaxies with stellar masses $8.5 < \log(M_*/M_{\sun}) < 10.0$. The survey covers the local Universe, $0.010 < z < 0.025$ ($= 40 \lta {\rm D/Mpc} \lta 110$), excluding the very nearby Universe to ensure that the $27''$ APEX beam recovers the total CO flux.  

Upon completion, the ALLSMOG survey will have CO measurements (both detections and upper limits) for $\sim100$ low mass ($<10^{10} \rm{M}_{\sun}$) galaxies. In this work, we use the `halfway' data release presented by \cite{2014MNRAS.445.2599B} which consists of 42 galaxies. Data observations and reduction were undertaken as described in that work. Being drawn from the SDSS spectroscopic survey, all ALLSMOG galaxies have available optical spectra. One of the aims motivating the ALLSMOG survey was to study the connection between metallicity and molecular gas content, and so all ALLSMOG galaxies have the complete suite of  optical line fluxes necessary for deriving metallicities (see \S\ref{sec:met-derive} for an explanation of the metallicity derivation). All ALLSMOG galaxies are selected from the SDSS spectroscopic survey, and have their stellar masses and star formation rates readily available. Both stellar masses and star formation rates, corrected for aperture effects caused by the fibre, are provided by the MPA-JHU group\footnote{http://www.mpa-garching.mpg.de/SDSS/DR7/}. Star formation rates are derived using optical line fitting, as described in \cite{2004MNRAS.351.1151B}. Stellar masses are likewise measured using the fitting technique described by \cite{2003MNRAS.341...54K}. Uncertainties on these parameters are typical for SDSS galaxies -- approximately $\sim 0.3$ dex and $\sim 0.15$ dex for SFRs and stellar masses, respectively.

The ALLSMOG galaxies used in this analysis have a metallicity range [$8.52 < 12+\log({\rm O/H}) <9.17$], a SFR range [$0.04 < {\rm SFR/M_{\sun}\,yr^{-1}} <12.0$], and lie between stellar masses [$8.5 < {\rm \log M_*/M_{\sun}} <10.0$].

All ALLSMOG galaxies published in the first data release are used in this analysis: a total of 42 galaxies.

\subsubsection{COLD GASS}

The COLD GASS survey is a large IRAM program, designed to survey molecular gas in massive $z\sim0$ galaxies, with $\log(M_*/M_{\sun}) >10 $. The survey covers the redshift range $0.025 < z < 0.05$ ($= 110 \lta {\rm D/Mpc} \lta 220$), and is selected at random from the parent GASS sample \citep{2010MNRAS.403..683C}. The COLD GASS sample is discussed in detail by \cite{2011MNRAS.415...32S}. 

As with the ALLSMOG sample, COLD GASS is drawn from the SDSS spectroscopic survey and has associated available spectra, which supply both stellar masses and star formation rates (as detailed in the ALLSMOG description above). We remove AGN (which contaminate the optical line fluxes, making the resultant metallically estimates inaccurate) from the COLD GASS sample using the BPT cut given by \cite{2003MNRAS.346.1055K}. The COLD GASS galaxies used here have a metallicity range [$8.87 < 12+\log({\rm O/H}) <9.29$], a SFR range [$0.04 < {\rm SFR/M_{\sun}\,yr^{-1}} <27$], and lie between stellar masses [$10.0 < {\rm \log M_*/M_{\sun}} <11.5$].

After discarding AGN, and galaxies with discrepancies between their R23 and [{\sc Nii}]/H$\alpha$ metallicities (see \S \ref{sec:met-derive} below), we are left with 115 COLD GASS galaxies.

\subsubsection{Herschel Reference Survey}

\begin{figure*}
\centering
\includegraphics[width=17cm]{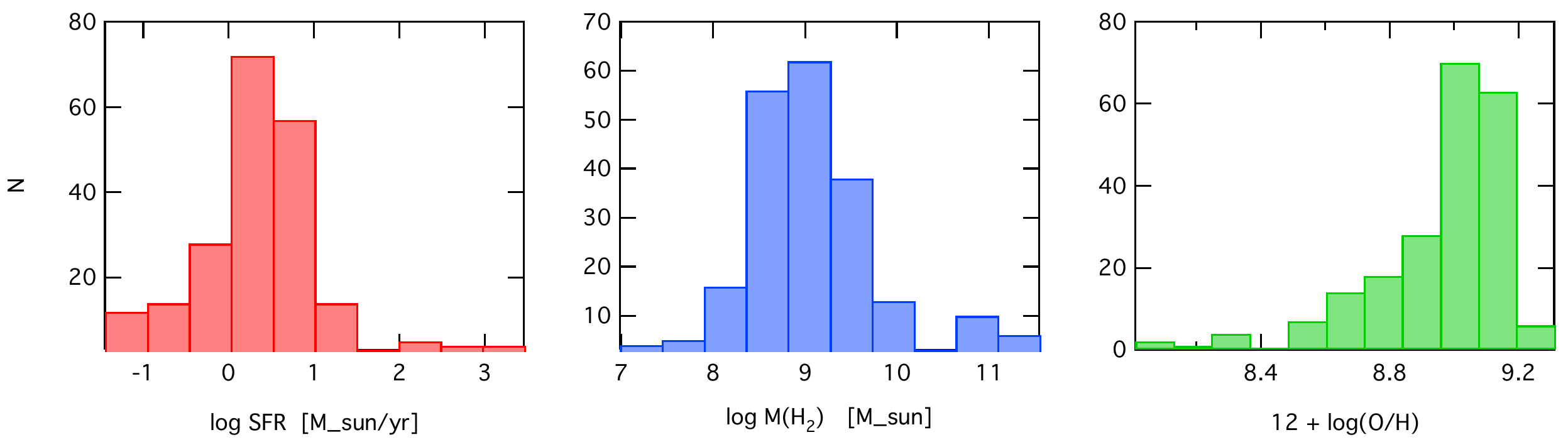}	
\caption{ Histogram of the star formation rates ({\it left panel}), molecular gas masses ({\it central panel}), and metallicities ({\it right panel}) of galaxies in the samples used in this work. As before, only galaxies that obey the selection criteria detailed in \S \ref{sec:sample} are shown.}
\label{fig:hist}
\end{figure*}

The Herschel Reference Survey (HRS) is a Herschel guaranteed time key project, designed to study a complete, volume limited sample of galaxies at distances $15 < {\rm D/Mpc} < 25$. Within these limits the survey is highly complete. A full survey description is given by \cite{2010PASP..122..261B}.

Lying closer than 25 Mpc, the HRS galaxies are too nearby to be effectively probed by a fibre survey such as SDSS (the $3''$ SDSS fibre would cover only a tiny fraction of the galaxy disc). However, ancillary parameters -- including both stellar masses and optical spectra -- are available from other sources. Stellar masses for the HRS are presented by \cite{2012A&A...544A.101C}, and were derived using $i$-band luminosities, with a typical uncertainty of $\sim 0.15$ dex. Star formation rates for the HRS galaxies can be calculated from a combination of UV and IR fluxes (UV fluxes presented by \citealt{2012A&A...544A.101C} and IR fluxes, presented by \citealt{2012MNRAS.423..197B}). Briefly, we define the FUV attenuation via the IRX' parameter,  IRX$ = \log({\rm L}_{\rm IR}/{\rm L}_{\rm FUV, obs})$:

\begin{equation}
{\rm A(FUV)} = -0.028X^3 + 0.392X^2 + 1.094X + 0.546,
\end{equation}

where $X$ is  `IRX'  \citep{2005MNRAS.360.1413B}. SFRs are then calculated from the attenuation-corrected FUV luminosity:

\begin{equation}
\log{\rm SFR} = \log {\rm L}_{\rm FUV, corr} - 9.68,
\end{equation}

which is a SFR prescription derived for use with the GALEX bands by \cite{2006ApJS..164...38I}. ${\rm FUV, corr} $ is the FUV luminosity, corrected for the effects of UV attenuation. We have reduced their original UV luminosity to SFR conversion factor by a factor of 1.5, in order to convert from their \cite{1955ApJ...121..161S} IMF to the \cite{2001MNRAS.322..231K} IMF used to calculate SDSS parameters (see \citealt{2007tS..173..267S} for details).

Optical spectra for the HRS, presented by \cite{2013A&A...550A.114B}, were taken by the 1.9m telescope at Observatoire de Haute Provence, and cover the spectral range 3600-6900\AA, allowing metallicities to be derived identically as for the SDSS samples described above. Again, we use a BPT cut to remove AGN from the sample, and do not use galaxies with discrepant R23 and [{\sc Nii}]/H$\alpha$ metallicities (as described by \S \ref{sec:met-derive} below). The HRS galaxies used here have a metallicity range [$8.57 < 12+\log({\rm O/H}) <9.05$], a SFR range [$0.70 < {\rm SFR/M_{\sun}\,yr^{-1}} <16.5$], and lie between stellar masses [$8.9 < {\rm \log M_*/M_{\sun}} <10.4$].

After these cuts, we are left with 24 HRS galaxies.

\subsubsection{Local Volume Legacy survey}
\label{sec:LVL}

In addition to the three dedicated CO surveys discussed above, we also include galaxies from the volume-limited Local Volume Legacy (LVL) survey. LVL is a IRAC and MIPS Legacy survey of a volume-limited sample of 258 galaxies, designed to be statistically complete within 11 Mpc. The LVL survey was designed to work in synergy with ancillary datasets, including the UV and H$\alpha$ survey 11HUGS (11 Mpc H$\alpha$ and UV Galaxy Survey). The combined dataset is volume-limited out to 11 Mpc (excluding the `zone of avoidanceÕ, defined by the Galactic plane $|b| \lta 20\degree$), with a magnitude limit $B<15$.

Stellar masses for the LVL sample have been estimated by \cite{Bothwell:2009aa}, by combining photometry with a colour-dependent mass to light ratio, taken from the models of \cite{2001ApJ...550..212B}. \cite{Bothwell:2009aa} quote the estimated uncertainty on the stellar masses as $\sim 0.3 $ dex. Gas data, both atomic and molecular, were compiled from a range of literature sources by \cite{Bothwell:2009aa}. Star formation rates for the LVL sample were derived by \cite{2009ApJ...706..599L}, by combining UV and IR fluxes as per the method outlined above. The average uncertainty on these SFRs is approximately $\sim 0.3$ dex.

The final required data product, optical metallicity data, is compiled and presented by \cite{2010ApJ...715..506M}. Requiring both molecular gas data and an optical metallicity measurement serves to eliminate much of the LVL sample: in total, 23 galaxies from LVL have the full complement of required data. (One of these galaxies, however -- UGC05364 -- has its metallicity estimated using observations of planetary nebulae -- a unique method, not shared by any other galaxy used in this analysis. We therefore discard UGC05364 from the sample.) The LVL galaxies in our combined sample have a metallicity range [$8.01 < 12+\log({\rm O/H}) <9.31$], a SFR range [$0.04 < {\rm SFR/M_{\sun}\,yr^{-1}} <5.2$], and lie between stellar masses [$8.4 < {\rm \log M_*/M_{\sun}} <11.5$].

The LVL sample used in this analysis consists of 22 galaxies.

\subsection{Galaxies at high redshift}


\subsubsection{Main sequence star-forming galaxies}

Our sample of high-$z$ `main sequence' star forming galaxies is taken from the PHIBBS survey for molecular gas in $1<z<3$ `main sequence' galaxies \citep{2013ApJ...768...74T}. These `main sequence' galaxies are selected via the `BX/BM' optical colour-selection criteria (which approximately selects `normal' star forming galaxies at high-$z$; details of this selection criteria are given in \citealt{2004ApJ...604..534S}). 

Galaxies in the PHIBBS survey lie at two distinct epochs:  $z \sim 1.2$ and $z \sim 2.2$. The lower redshift galaxies are unsuitable for inclusion in this work, as their optical spectra (provided by the DEEP2 survey) do not cover any suitable metallicity-tracing line ratios. The galaxies in the higher redshift bin, however, have been observed with VLT/SINFONI \citep{2012ApJ...761...43N}, and have publicly-available near-IR spectroscopy covering the H$\alpha$ and [NII] lines, which we use to derive gas-phase metallicities (as described in \S\ref{sec:met-derive} below).

Stellar masses, molecular gas data, and star formation rates for these galaxies have been taken from \cite{2013ApJ...768...74T}. To briefly summarise, stellar masses have been derived from UV-to-IR population synthesis modelling, and star formation rates were calculated using a combination of UV, optical continuum, H$\alpha$, and 24$\mu$m continuum fluxes. \cite{2013ApJ...768...74T} quote a typical stellar mass uncertainty of 0.13 dex, and a typical SFR uncertainty of 0.15 dex. Molecular gas data for this sample were obtained using the IRAM Plateau de Bure Interferometer. Due to the redshift of the sample, the sources here are observed in the $^{12}$CO($3-2$) line -- for the purposes of calculating gas masses, we have converted down to equivalent $^{12}$CO($1-0$) luminosities by assuming a `Milky Way'-like CO excitation, appropriate for these normal star-forming galaxies. The high-$z$ `main sequence' galaxies in our sample have a metallicity range [$8.61 < 12+\log({\rm O/H}) <9.05$], a SFR range [$30 < {\rm SFR/M_{\sun}\,yr^{-1}} <245$], and lie between stellar masses [$9.8 < {\rm \log M_*/M_{\sun}} <11.4$].

Our final sample of high-$z$ main sequence galaxies consists of 9 objects.

\subsubsection{Sub-mm galaxies (SMGs)}

We also include a sample of galaxies selected at sub-mm wavelengths (typically $S_{850} \simgt 5$mJy), known as sub-mm galaxies (SMGs). SMGs are high-redshift dusty galaxies with extreme SFRs ( $\simgt1000 {\rm \;M_{\sun}\, yr^{-1}}$). 
The source of the high SFRs exhibited by SMGs remains a matter of debate, but it is likely that the population contains both major mergers and extreme, `maximum starburst' disks. Either way, it is clear that SMGs represent an extreme population lying far above the star-forming `main sequence' at their redshifts. The inclusion of SMGs in our combined sample will extend the dynamic range of our analysis, in terms of both parameter space and the range of ISM environments sampled. 

Our sample of SMGs lie at  $z\sim2$, and are taken from the sample presented by \cite{2005ApJ...622..772C}. Metallicities were derived from Keck/NIRSPEC near-IR  spectroscopy \citep{2004ApJ...617...64S}, which covers both the H$\alpha$ and [N{\sc ii}] emission lines (again, see \S\ref{sec:met-derive} below for details of the metallicity derivation). We note that SMGs are highly dusty galaxies, and as such optical emission lines may not fully trace regions of active star formation. However in the absence of well-calibrated far-IR diagnostics (as described in, e.g., Nagao et al 2011), these optical lines remain the only way to estimate metallicities in SMGs. We therefore proceed with the metallicity calibration, and simply warn readers that the metallicity values for our SMGs may be uncertain. Stellar masses for these SMGs were derived from full population synthesis modelling by \cite{Hainline:2011lr}, and star formation rates were calculated using IR fluxes (derived from the far IR-radio correlation) by \cite{2013MNRAS.429.3047B}. Uncertainties on star formation rates and stellar masses for SMGs are typically of the order $0.3$ dex (as given by \citealt{Hainline:2011lr} and \citealt{2013MNRAS.429.3047B}).

 Molecular gas masses were taken from the CO survey by \cite{2013MNRAS.429.3047B}. As with the high-$z$ main sequence galaxies above, molecular gas masses are based on mid-$J$ $^{12}$CO measurements -- either $^{12}$CO($4-3$) or $^{12}$CO($3-2$). We have adjusted these down to the equivalent $^{12}$CO($1-0$) luminosities using the SMG CO excitation derived by \cite{2013MNRAS.429.3047B}. One SMG (SMMJ030227.73+000653.3) has an unusually high [N{\sc ii}]/H$\alpha$ ratio; as noted by \cite{2004ApJ...617...64S}, this is likely indicative of contamination by AGN activity. As such, we exclude it from our analysis. 

The SMGs  in our analysis have a metallicity range [$8.74 < 12+\log({\rm O/H}) <9.09$], a SFR range [$390 < {\rm SFR/M_{\sun}\,yr^{-1}} <2890$], and lie between stellar masses [$10.3 < {\rm \log M_*/M_{\sun}} <11.2$].

Excluding SMMJ030227.73+000653.3, our final SMG sample consists of 9 galaxies.

\subsection{Deriving physical parameters}

\subsubsection{Deriving metallicities}
\label{sec:met-derive}

For all galaxies in this work (with the exception of some galaxies in the LVL sample: see below), we derive gas-phase metallicities using optical strong-line fluxes. Where possible, we take optical line fluxes from SDSS (this is applicable for the ALLSMOG and COLD GASS samples). Following Mannucci et al (2010), we selected galaxies for inclusion in our ALLSMOG sample by requiring a signal-to-noise ratio of 25 in the H$\alpha$ line -- this ensures that the main optical lines are generally detected at high enough S/N to reliably derive a metallicity. All galaxies in the COLD GASS sample also fulfil this criteria. Mean line flux errors for our SDSS sources are as follows: H$\alpha_{\rm err} \sim 4\%$; [NII]$_{\rm err} \sim 5\%$; H$\beta_{\rm err} \sim 25\%$; [OII]3727$_{\rm err} \sim 48\%$; [OIII]5007$_{\rm err} \sim 28\%$; [OIII]4959$_{\rm err} \sim 31\%$. The faintest line detected is generally [OII]3727, and on average sources in our samples have [OII]3727 detected with a S/N ratio of $\sim 2$. All lines are individually corrected for dust extinction, which is estimated using the  $H\alpha/H\beta$ ratio, and a Milky Way extinction curve.

For ease of comparison, we follow  \cite{2010MNRAS.408.2115M} in estimating metallicity based on two independent diagnostics: (1) the N[{\sc ii}]/H$\alpha$ ratio, and (2) the R23 parameter, $= ({\rm [OII]3727 \, + \, \rm [OIII]4958,5007}) / H\beta$.  We then use the abundance calibrations of \cite{2008A&A...488..463M} to calculate two separate metallicity measurements, one for each diagnostic. We take as our final metallicity value the mean of these two metallicity measurements, discarding any galaxies for which the methods give answers discrepant by $>0.2$ dex: such galaxies are likely to have inaccuracies in one or both metallicity diagnostics, and as such are unreliable. We note that 9 galaxies in our ALLSMOG sample lie at $z < 0.019$, and thus the [OII]3727 line does not fall within the SDSS spectral range. For these 9 galaxies, instead of the R23 parameter (which requires the detection of the [OII]3727 line) we adopt the metallicity as listed in the MPA-JHU catalogue (which is derived following Tremonti et al. 2004), adjusted for our metallicity calibration. We then, as above, take the mean of this value and the value derived using the N[{\sc ii}]/H$\alpha$ ratio.

One potential source of metallicity bias in our SDSS-observed samples is due to aperture effects. SDSS is a fibre-based spectroscopic survey, and the 3$''$ fibre may only sample the central few kpc of some galaxies, if they are nearby, and subtend large areas on the sky. Given the existence of metallicity gradients, this may cause abundance estimates to be biased high (though this effect is likely to be small, as gradients in star-forming galaxies tend to be shallow - $\sim 0.1$ dex/R$_{\rm e}$; \citealt{2014A&A...563A..49S}). \cite{2005PASP..117..227K} estimate that a `fibre covering fraction' of 20\% is sufficient to recover the metallicity of a galaxy without aperture bias. If we remove the ALLSMOG and COLD GASS galaxies which fail this criteria (3 ALLSMOG galaxies, and 33 COLD GASS galaxies), our results are unaffected. For the remainder of this work, we therefore proceed with the full samples as described above.

To recap the various sources of the non-SDSS spectroscopy, HRS galaxies have long-slit integrated spectroscopy presented by \cite{2013A&A...550A.114B}, high-$z$ `main sequence' galaxies have near-IR spectra presented by \cite{2012ApJ...761...43N}, and the SMGs have near-IR spectra presented by \cite{2004ApJ...617...64S}. For the galaxies at high-redshift (both main-sequence and SMGs), the full suite of optical strong lines is not available for each source. Instead, we estimate their metallicities based on the N[{\sc ii}]/H$\alpha$ ratio alone, again using the abundance calibrations provided by  \cite{2008A&A...488..463M}.

As mentioned in \S\ref{sec:LVL} above, galaxies in the LVL sample have metallicities compiled from the literature by \cite{2010ApJ...715..506M}. As noted by those authors, the metallicity data for many of these galaxies has been amassed over a number of years from a large number of disparate observations. As such it is challenging to put all the LVL metallicities into a common calibration framework. Indeed, \cite{2010ApJ...715..506M} do not attempt to do so, simply noting that the non-uniform calibrations add an additional uncertainty across the sample of $\sim 0.2$ dex. We follow this approach, and do not attempt to standardise the LVL metallicity measurements. Instead, we apply an additional 0.2 dex of uncertainty to all LVL metallicity measurements when fitting to the data. We stress that none of our conclusions are dependent on the inclusion of LVL -- repeating our analysis without the LVL sample does not significantly change our results. 


\subsubsection{Deriving H$_2$ masses}

Molecular hydrogen masses can be calculated from the CO luminosity, $L'_{\rm CO}$, by assuming a CO/H$_2$ conversion factor $\alpha_{\rm CO}$:

\begin{equation}
{\rm M(H}_2) = \alpha_{\rm CO}L'_{\rm CO(1-0)}
\end{equation}

There has been a vast amount of work over the last few decades dedicated to empirically measuring, and theoretically modelling, $\alpha_{\rm CO}$ (\citealt{2013ARA&A..51..207B} gives a recent review). While a number of physical factors can cause $\alpha_{\rm CO}$ to vary, it is likely that the dominant factor is metallicity, with $\alpha_{\rm CO}$ increasing rapidly as metallicity decreases. In this work, we will present results by adopting a CO/H$_2$ conversion factor that varies with metallicity. Several metallicity-dependent CO/H$_2$ conversion factor prescriptions are given in a recent review by \cite{2013ARA&A..51..207B}. In particular, we will explore results derived using four recent metallicity-dependent conversion factors, derived by \cite{2010ApJ...716.1191W}, \cite{2011MNRAS.412..337G}, \cite{2012MNRAS.421.3127N}, and \cite{2012ApJ...747..124F}.

Uncertainties on molecular gas masses are dominated by systematic uncertainty concerning the CO/H$_2$ conversion factor. Uncertainties on CO fluxes, absolute flux calibration, and aperture corrections contribute, at most, $\sim 20\%$ uncertainty to the final value of M(H$_2$). The systematic uncertainty due to the CO/H$_2$ conversion factor is difficult to estimate (\citealt{2011MNRAS.415...32S} estimate a combined uncertainty, including systematic uncertainty on the conversion factor, of 0.3 dex). We have attempted to mitigate systematics due to the conversion factor by calculating our results for a range of metallicity-dependent factors. For the purposes of our analysis, however, we adopt a conservative total uncertainty of 0.3 dex on our molecular gas masses.




\section{Results: Molecular gas and the mass-metallicity relation}


\begin{figure}
\centering
\mbox
{
\includegraphics[width=8cm, clip=true, trim=50 350 200 150]{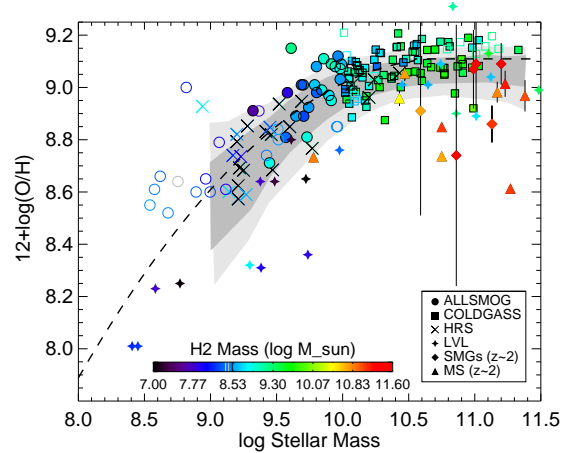}
}
\caption{The mass-metallicity relation for the galaxies in our sample. The grey shaded areas are taken from Tremonti et al. (2004), and show the area that contains 64\% (light shaded area) and 90\% (dark shaded area) of all SDSS galaxies. We have adjusted the metallicity scaling for the Maiolino et al. (2008) calibration used in this work. We have calculated molecular gas masses using the (metallicity-dependent) CO/H$_2$ conversion factor from Wolfire et al. (2010). The dashed line is the mass-metallicity fit to the sample of galaxies used in this work. Galaxies at high-$z$, which have greater uncertainty on their metallicity estimates, are shown with error bars. Open symbols show galaxies not detected in $^{12}$CO, at the position of the 3$\sigma$ upper limits on their gas mass.}
\label{fig:MZR}
\end{figure}
\begin{figure*}
\centering
\mbox
{
\includegraphics[width=14cm]{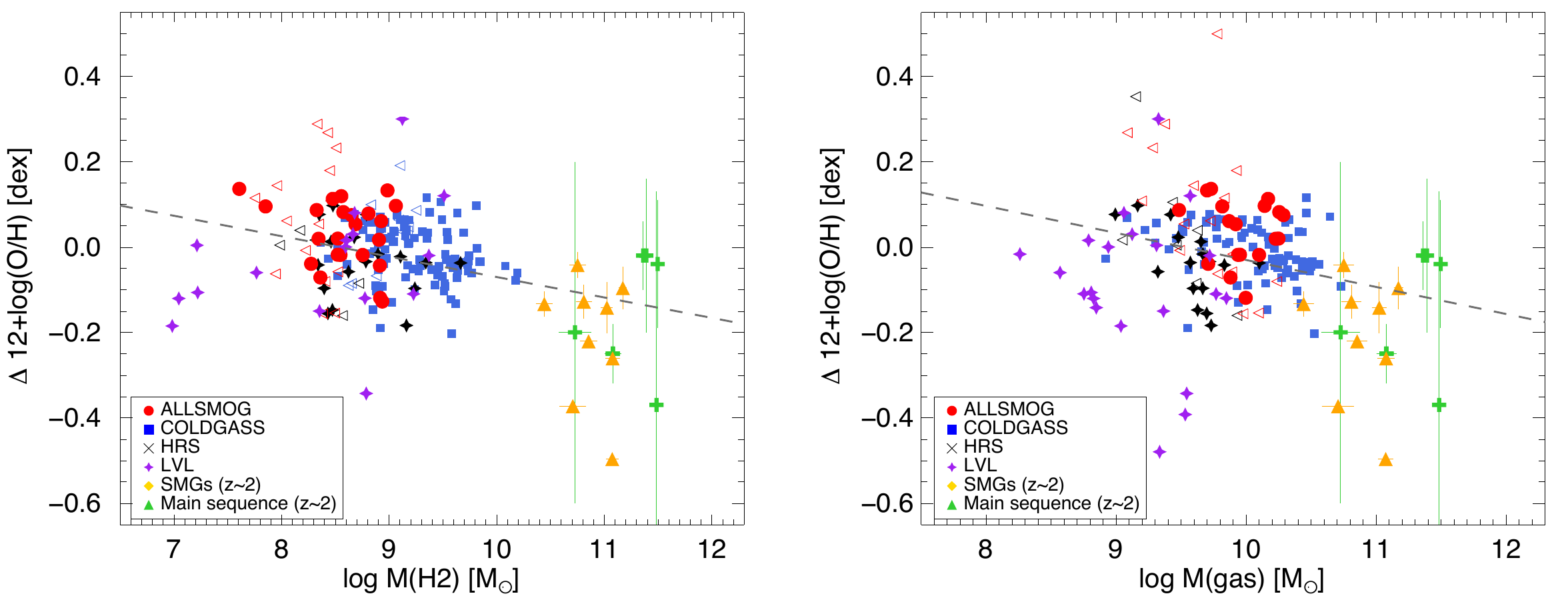}	
}
\caption{The offset from the $z=0$ mass-metallicity relation, plotting against molecular hydrogen mass ({\it left panel}), and total gas mass ({\it right panel}). In each plot, a linear fit is over-plotted. It can be seen that there is an inverse correlation between mass-metallicity offset and both H$_2$ mass and total gas mass (so galaxies with lower than average gas masses lie systematically above the mass-metallicity relation).}
\label{fig:dz_mh2}
\end{figure*}

We turn now to an examination of the effect of molecular gas on the scatter in the mass-metallicity relation . As discussed in the introduction,  significant relationships have been discovered between the mass-metallicity relation  and both SFR (the `FMR'; \citealt{2010MNRAS.408.2115M}; \citealt{2010A&A...521L..53L}) and \HI\ \citep{2013MNRAS.433.1425B}. Any relationship with molecular gas has thus far proved elusive, due to a likely combination of small available sample sizes, and potential degeneracy between metallicity and molecular gas masses (derived using a conversion factor which is itself strongly dependent on metallicity). The combined sample assembled in this work represents a uniquely ideal dataset for examining any potential trends between metallicity and molecular gas mass, due to its dynamic range in both parameters.

Figure \ref{fig:MZR} shows the mass-metallicity relation for galaxies in our combined sample. The original samples are differentiated with different symbols (shown in the inset key). The combined sample broadly follows the established $z\sim0$ mass-metallicity relation (the dark and light shaded regions in the background show the 1$\sigma$ and 2$\sigma$ distributions in the mass-metallicity relation, as measured for SDSS galaxies by \citealt{2004ApJ...613..898T}, adjusted for the metallicity calibrations used in this work). As expected, higher-$z$ galaxies generally fall below the $z\sim0$ mass-metallicity relation (we discuss this below). Most individual samples follow the mass-metallicity relation reasonably closely: several ALLSMOG galaxies, not detected in CO emission lie above the relation, while some LVL galaxies lie 0.2-0.5 dex below the relation.

The symbol colour-coding in Fig. \ref{fig:MZR} shows the molecular gas mass, as derived using the Wolfire et al. (2010) CO/H$_2$ conversion factor. A molecular gas `FMR' effect would be apparent if galaxies at a constant stellar mass have an inverse relation between molecular gas mass and metallicity
\footnote{Note that we refer to the `FMR' effect as being an inverse relation with metallicity at a given stellar mass; due to the low numbers of galaxies available with molecular gas measurements, we are unable to explore the potential redshift independence of our relations, another hallmark of the FMR.}. 
Such a relation is difficult to discern by eye, due to both the uncertainty-driven scatter, and the the strong relation between molecular gas mass and stellar mass. To make any potential trends clearer, we examine the relation between molecular gas mass and the {\it offset} from the mass-metallicity relation (which we calculate by performing a quadratic fit to the mass-metallicity relation defined by our combined sample). The quadratic fit used for calculating offsets is shown as a dashed line in Fig. \ref{fig:MZR} (note that we have flattened the fit at upper end, to maximise at the `saturation metallicity', 12+log(O/H)=9.1). The fit is in good agreement with the fit found for a far larger SDSS sample by \cite{2004ApJ...613..898T}, but is shifted slightly towards higher metallicities at stellar masses $\log {\rm M_*} > 10^{10} {\rm M_{\sun}}$ .

Figure \ref{fig:dz_mh2} (left panel) shows the offset from the mass-metallicity relation, plotted against molecular gas mass. For legibility, we have only plotted error bars for the high-$z$ galaxies, which have their metallicities measured using the [NII]/H$\alpha$ calibrator alone and as a result are more uncertain (due to the greater flux errors on the more distant objects, as well as the uncertainty due to the availability of just a single calibrator).


\begin{figure*}
\centering
\includegraphics[width=14cm]{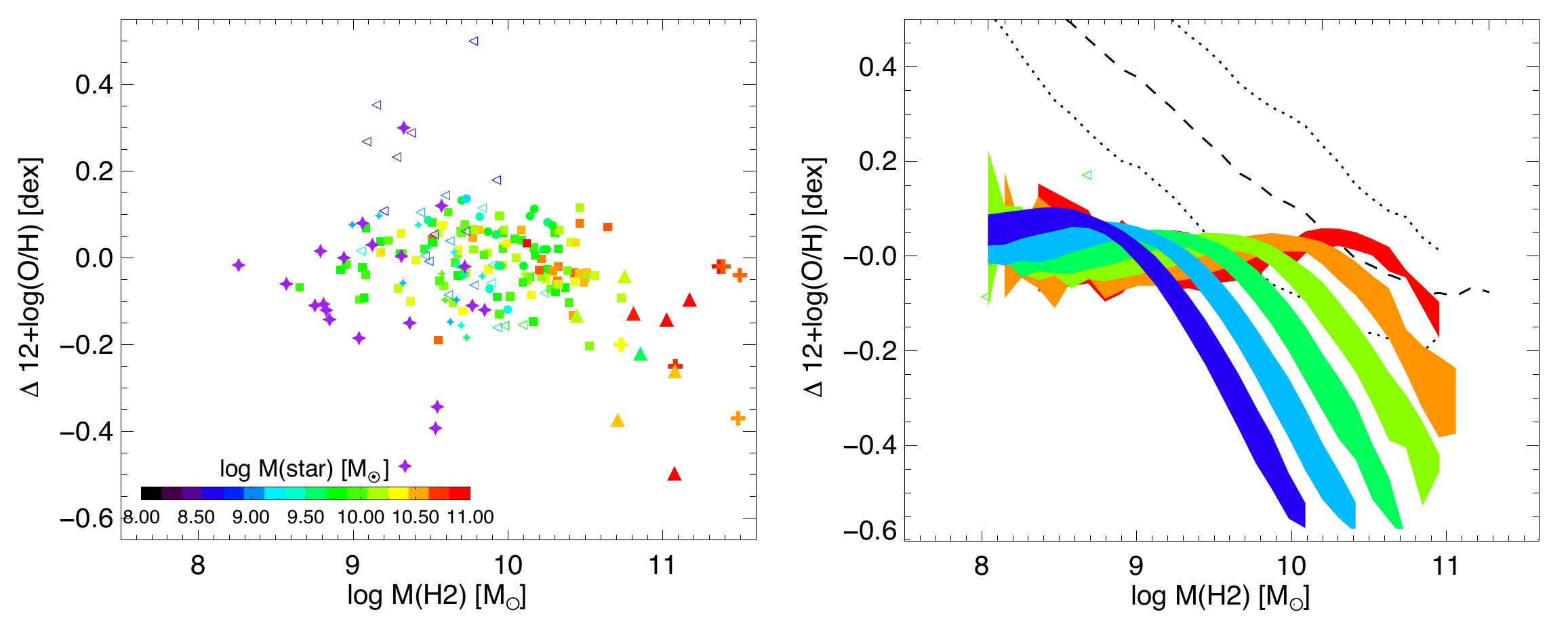}
\caption{The offset from the $z=0$ mass-metallicity relation, plotted against total gas mass, for (left panel) our observed sample, and (right panel) calculated using the Peng \& Maiolino (2014) gas regulator model. Model data are shown in bins of stellar mass of 0.5 dex width. The model predicts a strong inverse dependence between mass-metallicity offset within each stellar mass bin. Also shown is the predicted `quenching sequence' (dashed line, with associated standard deviation indicated with dotted lines).}
\label{fig:model_overlay}
\end{figure*}

We find that there is a shallow, but clear, inverse dependence between molecular gas mass and the offset from the mass-metallicity relation. Fitting linear functions to the relation, we find a slope of $= -0.088 \pm 0.021$ (for the  \cite{2010ApJ...716.1191W}  conversion factor;  slopes derived using our other conversion factors are: {\cite{2011MNRAS.412..337G}  $= -0.078 \pm 0.019$; \cite{2012MNRAS.421.3127N} $ = -0.098 \pm 0.020$; and \cite{2012ApJ...747..124F} = $ -0.081 \pm 0.018$). There is therefore clear evidence for a molecular gas `FMR' effect, whereby galaxies at a fixed stellar mass show an inverse dependence between their metallicity and their molecular gas mass. 

Figure \ref{fig:dz_mh2} (right panel) shows the offset from the mass-metallicity relation, plotted against total (i.e., H$_2$+\HI) gas mass. For our high-$z$ galaxies, we have assumed that their ISM is entirely molecular (i.e., their total gas mass is equivalent to their molecular gas mass). As for the molecular gas above, we find that for each of four metallicity-dependent conversion factors there is a shallow inverse dependence between total gas mass and the offset from the mass-metallicity relation. Given the dependence with molecular gas demonstrated above, and the dependence with atomic gas presented by Bothwell et al. (2013), a dependence with total gas mass is of course unsurprising.  Fitting linear functions the relation, we find a slope of $= -0.076 \pm 0.022$ (again, this is for the \cite{2010ApJ...716.1191W}  conversion factor; other slopes are:  \cite{2011MNRAS.412..337G}  $= -0.071 \pm 0.016$; \cite{2012MNRAS.421.3127N} $ = -0.083 \pm 0.018$; and \cite{2012ApJ...747..124F} = $ -0.077 \pm 0.017$). There is therefore also evidence for a total gas `FMR' effect, whereby galaxies at a fixed stellar mass show an inverse dependence between their metallicity and their total (i.e., H$_2$+\HI) gas mass. 

\subsection{Comparison to an analytical model}

In Fig. \ref{fig:model_overlay} we have compared the right panel of Fig. \ref{fig:dz_mh2} to the prediction from the analytical `gas regulator' model of Peng \& Maiolino (2014). We have binned the model data ($\sim 5\times 10^5$ model galaxies) in stellar mass bins of width 0.5 dex, and colour-coded both the model tracks and the observational data by stellar mass.

The model makes two main predictions. Firstly, the model predicts that within a stellar mass bin, the offset from the mass-metallicity relation does increase (in the negative direction, i.e., below the mass-metallicity relation) with increasing total gas mass. This is broadly the same trend as we have found for our observational data above. However, the model also demonstrates that the dependency of the mass-metallicity offset is far steeper within any individual stellar mass bin than it would be for the sample as a whole (due to the strong correlation between stellar mass and total gas mass). However, such a separation in stellar mass is not immediately clear in the observational data -- there seems to be sufficient scatter in the gas mass-stellar mass correlation that any potentially clear trends are obscured. The model does demonstrate why, when considering the the sample as a whole, we only see a shallow dependence between metallicity offset and gas mass.



\section{RESULTS: Principle Component Analysis}
\label{sec:PCA}

\begin{figure*}
\centering
\mbox
{
\includegraphics[clip=true, trim=50 370 80 50, width=16cm]{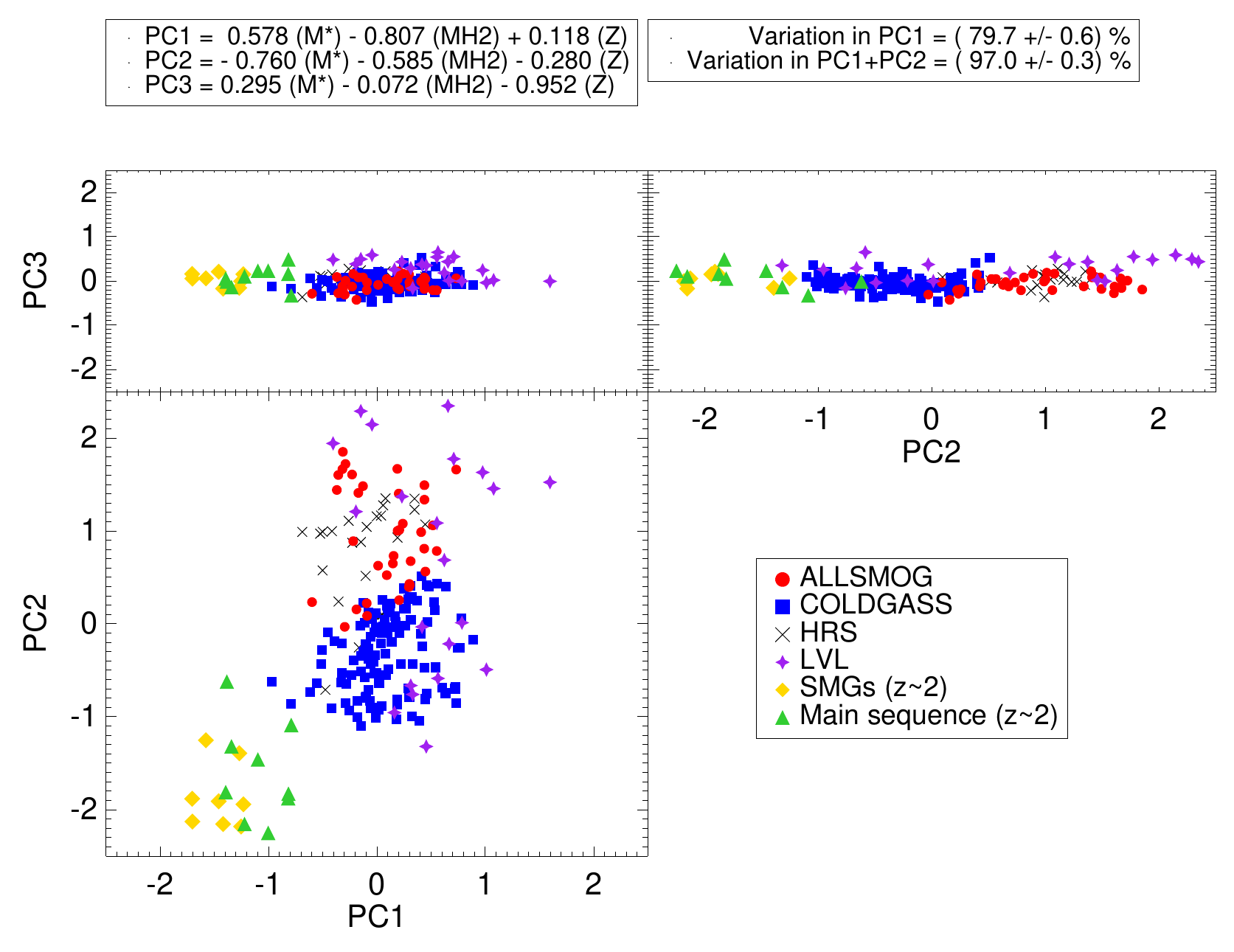}	
}
\caption{Three projections of the 3-space defined by the principle components derived in \S \ref{sec:PCA}. The inset legends above the plots give the definitions of the principle components, in terms of stellar mass (M$*$), molecular gas mass (MH2), and metallicity (Z).}
\label{fig:vary_Mh2}
\end{figure*}

\begin{figure*}
\centering
\mbox
{
\includegraphics[clip=true, trim=50 370 80 50, width=16cm]{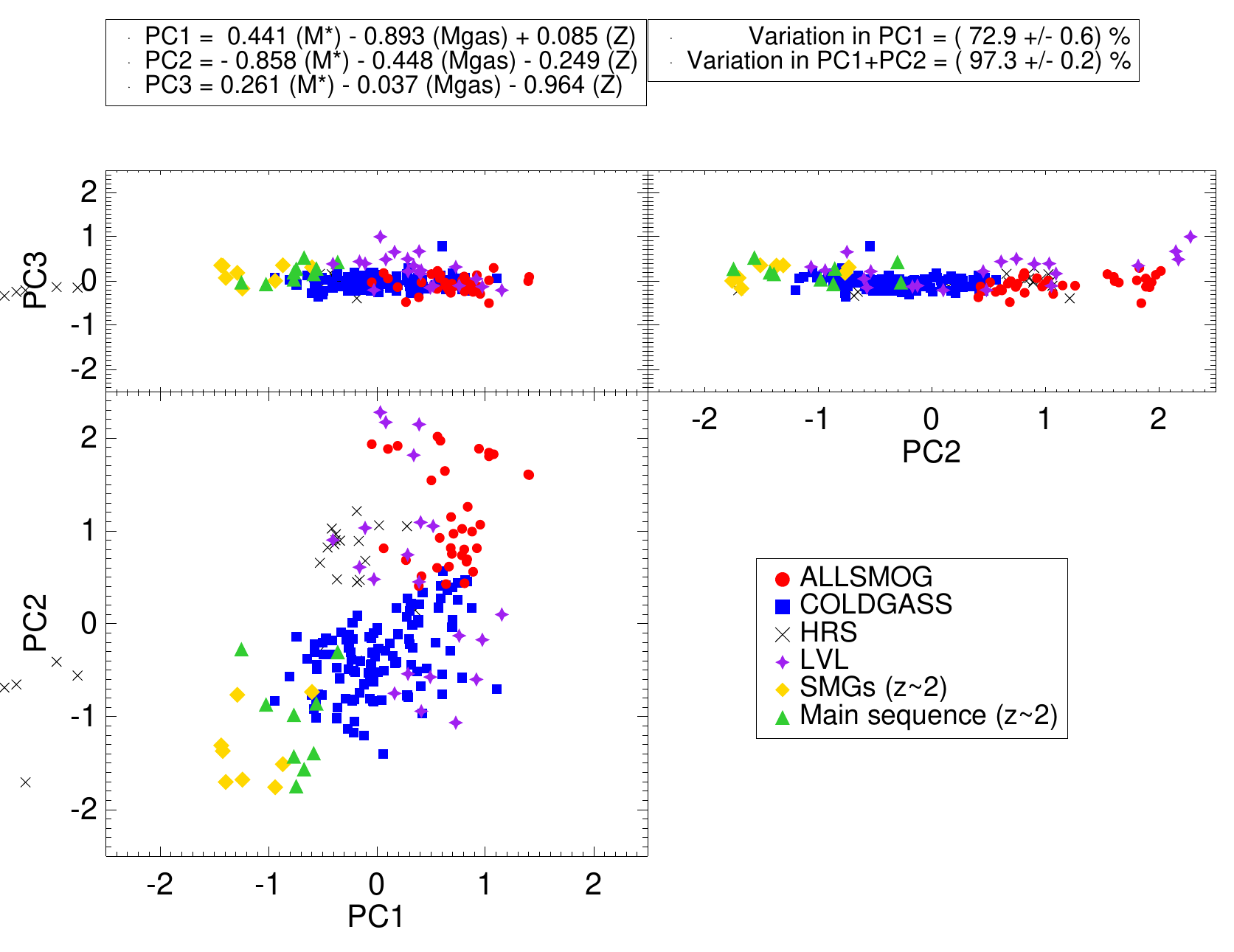}	
}
\caption{Three projections of the 3-space defined by the principle components derived in \S  \ref{sec:PCA}. The inset legends above the plots give the definitions of the principle components, in terms of stellar mass (M$*$), total gas mass (Mgas), and metallicity (Z).}
\label{fig:vary_Mgas}
\end{figure*}


\begin{figure*}
\centering
\mbox
{
\includegraphics[clip=true, trim=50 370 80 50, width=16cm]{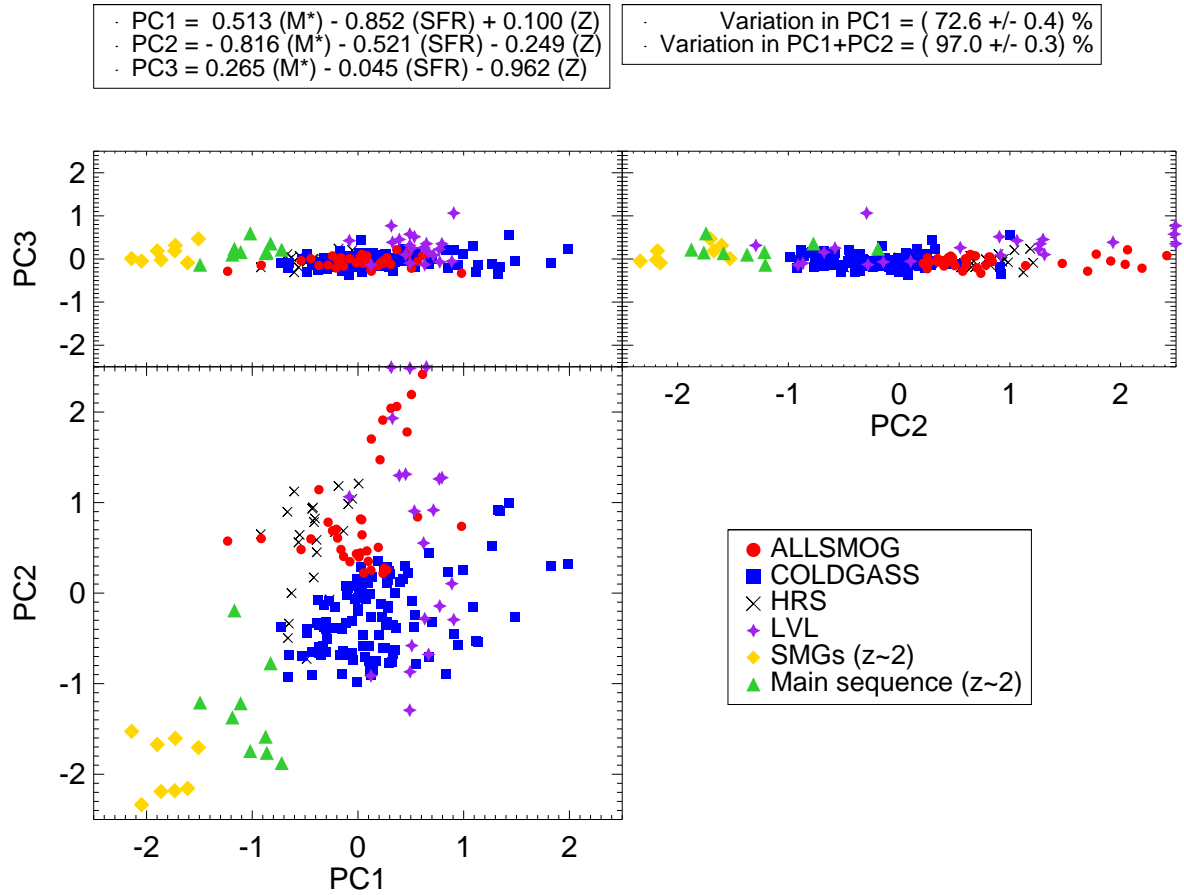}	
}
\caption{Three projections of the 3-space defined by the principle components derived in \S  \ref{sec:PCA}. The inset legends above the plots give the definitions of the principle components, in terms of stellar mass (M$*$), star formation rate (SFR), and metallicity (Z).}
\label{fig:vary_SFR}
\end{figure*}


\begin{figure*}
\centering
\mbox
{
\includegraphics[clip=true, trim=50 370 80 50, width=16cm]{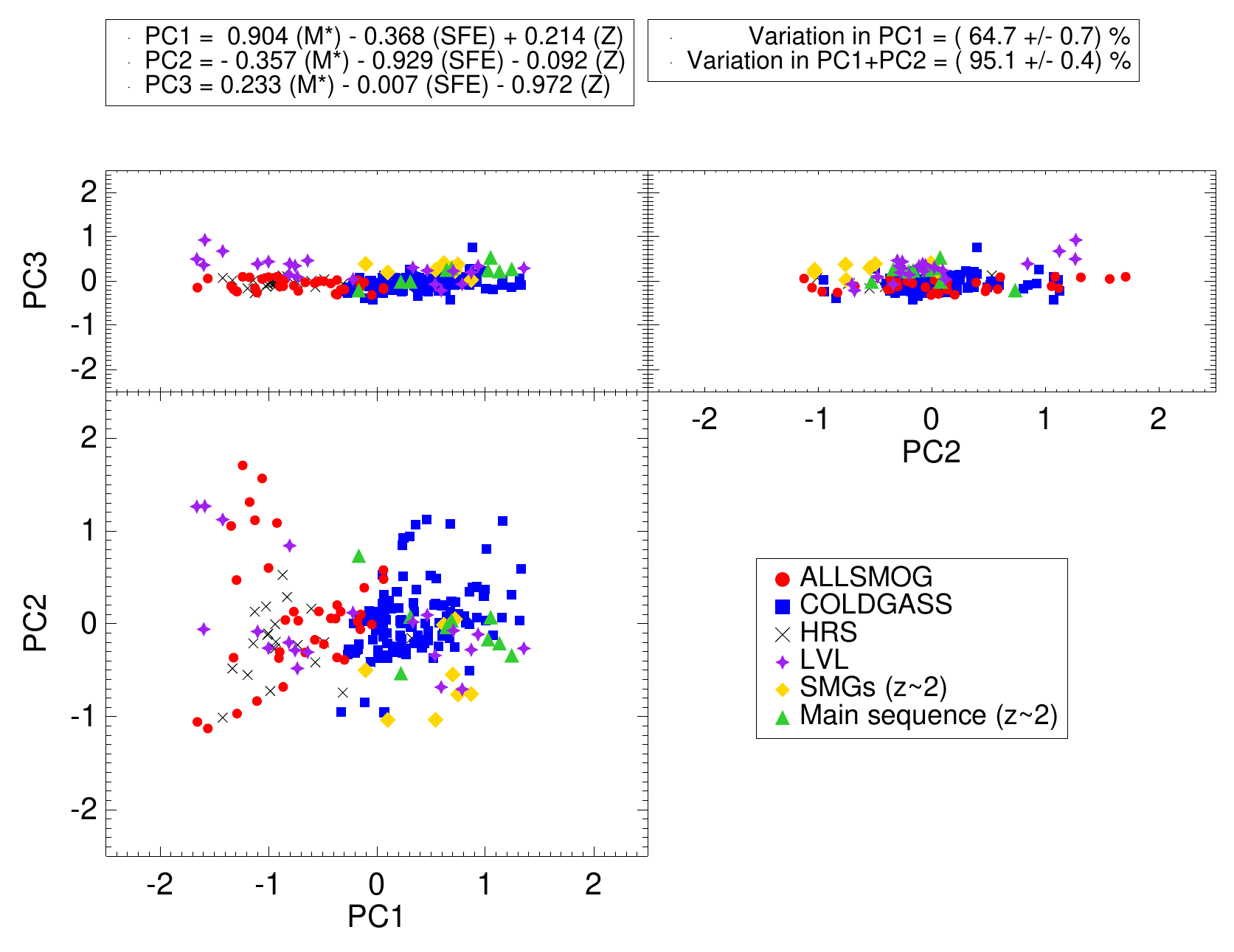}	
}
\caption{Three projections of the 3-space defined by the principle components derived in \S \ref{sec:PCA}. The inset legends above the plots give the definitions of the principle components, in terms of stellar mass (M$*$), star formation efficiency (SFE), and metallicity (Z). Metallicity has a weaker effect on the distribution, which can be mainly defined in terms of M$*$ and SFE.}
\label{fig:vary_SFE}
\end{figure*}


The analysis contained within the previous sections suffers from two potential weaknesses. Firstly, as previously discussed, the correlation between gas mass and stellar mass serves to erode any clear trends between mass-metallicity offset and gas mass; other than showing that the analytical model is capable of reproducing the observed data, it is challenging to quantify any trends. Secondly, the technique suffered from being parametric -- that is, dependent on the assumption that some distribution exists between stellar mass and metallicity, and paramerising data points in terms of their offset from a quadratic fit to that relation. In our case, the specific results obtained depend strongly on the form of the fit to the mass-metallicity relation. Ideally, we would like to frame the analysis in terms of a non-parametric fitting technique, which simply examines the correlations in the data without presupposing any specific relationship between them. One such technique is Principle Component Analysis (PCA).
  
Principle Component Analysis (PCA) is a linear transformation technique, by which a set of physical variables are converted into a set of orthogonal, linearly uncorrelated components (called `Principle Components') defined by a set of eigenvectors. The conversion is chosen so to ensure that the maximum possible variance in the data is contained within the first component, and each subsequent component contains as much variance as possible (with the constraint that each component is orientated orthogonally to every preceding component). In practice, PCA performs a coordinate transformation revealing the optimum `projection' of a dataset, and revealing which parameters are responsible for the variance in the sample. PCA is particularly useful for revealing any possible `dimensionality reduction' -- for example, revealing that a distribution forms a 2D `plane' in 3D parameter space (as \citealt{2012MNRAS.427..906H} found for stellar mass, SFR and metallicity). PCA was used by both \cite{2010A&A...521L..53L} and \cite{2012MNRAS.427..906H} to examine the secondary dependence of SFR on the mass-metallicity relation. 


Here, we perform PCA on our data, examining the stellar mass, metallicity, and a third parameter of interest. In turn, we will examine the molecular gas mass, the {\it total} gas mass, the star formation rate, and the star formation efficiency. We first normalise each parameter to the mean value for our combined sample:

$$ \log {\rm (M_*)^{PCA}}=      \log {\rm (M_*)} - 10.13$$
$${\rm 12+log(O/H)^{PCA}}=   {\rm 12+log(O/H)}  -  8.95$$
$$\log {\rm (SFE)^{PCA}}= \log {\rm (SFE)}  + 8.68$$
$$\log {\rm (SFR)^{PCA}}=  \log ({\rm SFR}) -    0.39$$
$$\log {\rm (M_{H2})^{PCA}}= \log {\rm (M_{H2})} -     9.07$$
$$\log {\rm (M_{gas})^{PCA}} = \log {\rm (M_{gas})}    -  9.71$$

We have accounted for uncertainty by adopting a Monte-Carlo approach, performing $10^5$ PCA iterations for each combination of parameters. For each iteration, each galaxy has the value of its physical parameters randomly perturbed by an amount following the respective error on each parameter. At the end of the $10^5$ iterations, we take the mean and standard deviation of the resulting Eigenvector distribution to be our final Eigenvector values (and uncertainties thereon). In order to ensure we are not unduly influenced by outliers, we also perform sample bootstrapping during the Monte-Carlo: for each iteration, we randomly sample (with replacement) our complete sample of galaxies, generating in each case a sample size equal to our original dataset.

It must be noted that a potential weakness in the application of PCA to our data is that PCA can only describe datasets in terms of linear relationships between parameters -- more complex distributions of data cannot be described in terms of a simple set of orthogonal eigenvectors. While linear relationships (such as the SFR-M$_*$ `main sequence') are ideal for modelling using PCA, applying the technique to non-linear correlations (such as the mass-metallicity relation) will, be definition, be somewhat inaccurate. In practice, this will have the effect of increasing the apparent scatter around the component vectors. Given the relatively low numbers of galaxies in our sample, it is likely that the uncertainty added by forcing linear relations onto the data is not larger than the uncertainty inherent in the distribution of the data (as revealed by our Monte-Carlo approach). We therefore continue to perform PCA on our data, but caution the reader that some uncertainty may be added due to non-linearity in some underlying correlations.

The results are discussed below. Fig. \ref{fig:vary_Mh2} shows the results of performing PCA on the parameters stellar mass, metallicity, and molecular gas mass. Appendix Figures \ref{fig:vary_Mgas}, \ref{fig:vary_SFR}, and \ref{fig:vary_SFE} show the results when using total gas mass, star formation rate, and star formation efficiency (respectively) as the third parameter. Table \ref{tab:PCA} lists the values of the Eigenvectors of the components in each instance. 

In each of the four cases, the PCA results do indeed suggest that the galaxy form a 2D plane in the parameter space, with just $\sim1.5\% - 3\%$ of the total variation occurring in the third principle component. As noted by  \cite{2012MNRAS.427..906H}, with very little of the sample variance contained within the third principle component, it is a useful tool for examining the optimal relationship between the three parameters of interest. In each of our cases, the third parameter is highly dominated by metallicity, with only minor contributions from the other parameters. We can therefore `solve' for metallicity, by setting the third principle component equal to zero (which is valid at the $\sim1.5\% - 3\%$ level), allowing us to examine the effect of each parameter (stellar mass, molecular gas mass, total gas mass, SFR, and SFE) on metallicity. 

\begin{flushleft}
\textbf{Molecular gas mass}:
\end{flushleft}

The PCA results for molecular gas are shown in Fig. \ref{fig:vary_Mh2}. $\sim 80\%$ of the variance is contained within the first component, and $\sim 97\%$ of the variance is contained within the first two. The first component is defined mainly in terms of stellar mass and molecular gas mass (molecular gas mass being the dominant contributor), with a small contribution from metallicity. The second component is similarly defined, with the contribution from metallicity being a little greater. The third component, which accounts for just $2.8\%$ of the sample variation, is dominated by metallicity (telling us that that metallicity has the minority contribution to the total sample variance). 

We now set the third principle component to zero. Here, the third principle component is defined as 

\begin{multline}
{\rm PC 3} = 0.295 \;{\rm (M_*)^{PCA}} - 0.072 \;{\rm (M_{H2})^{PCA}} \\ -0.952 \; {\rm 12+log(O/H)^{PCA}}
\end{multline}

Setting this to zero implies that:

\begin{equation}
{\rm 12+log(O/H)} = 0.31 {\rm (log \; M_*)} - 0.08 {\rm (log \; M_{H2}) + 6.53}
\end{equation}

That is, the metallicity is primarily determined by the stellar mass (i.e., the mass-metallicity relation), with a  secondary dependence on the molecular gas mass. This secondary dependence on molecular gas mass is approximately 22\% times as strong as the primary dependence on stellar mass.

\begin{flushleft}
\textbf{Total gas mass}:
\end{flushleft}

The PCA results for total gas mass are shown in Fig. \ref{fig:vary_Mgas}. $\sim 73\%$ of the variance is contained within the first component, and $\sim 98\%$ of the variance is contained within the first two. The first component is defined mainly in terms of stellar mass and total gas mass, with a small contribution from metallicity. And again, the second component is similarly defined, with the contribution from metallicity being a little greater. The third component, which here accounts for just $2.5\%$ of the sample variation, is dominated by metallicity. 

Again, we set the third principle component to zero. Here, the third principle component is defined as 

\begin{multline}
{\rm PC 3} = 0.261 \;{\rm (M_*)^{PCA}} - 0.038 \;{\rm (M_{gas})^{PCA}} \\ -0.964 \; {\rm 12+log(O/H)^{PCA}}
\end{multline}

Setting this to zero implies that:

\begin{equation}
{\rm 12+log(O/H)} = 0.27 {\rm (log \; M_*)} - 0.04 {\rm (log \; M_{gas}) + 6.60}
\end{equation}

That is, the metallicity is primarily determined by the stellar mass, with a secondary dependence on the total gas mass. This secondary dependence on total gas mass is approximately 13\% times as strong as the primary dependence on stellar mass. This shows that the metallicity dependence with the molecular gas content (discussed in the previous sub-section) is stronger than with the total gas content.

\begin{flushleft}
\textbf{Star formation rate}:
\end{flushleft}

The PCA results for SFR are shown in Fig. \ref{fig:vary_SFR}. $\sim 73\%$ of the variance is contained within the first component, and $\sim 97\%$ of the variance is contained within the first two. The first component is defined mainly in terms of stellar mass and SFR, with a small contribution from metallicity. And again, the second component is similarly defined, with the contribution from metallicity being a little greater. The third component, which here accounts for just $2.8\%$ of the sample variation, is dominated by metallicity. These results are consistent with the work done by \cite{2012MNRAS.427..906H}, who used PCA to examine correlations between stellar mass, SFR and metallicity for a sample of metal-poor starbursts. 

Setting the third principle component to zero. Here, the third principle component is defined as 

\begin{multline}
{\rm PC 3} = 0.265 \;{\rm (M_*)^{PCA}} - 0.045 \;{\rm (SFR)^{PCA}} \\ -0.963 \; {\rm 12+log(O/H)^{PCA}}
\end{multline}

Setting this to zero implies that:

\begin{equation}
{\rm 12+log(O/H)} = 0.28 {\; \rm (\log\, M_*)} - 0.05 {\rm (\log \, SFR) + 6.14}
\end{equation}

That is, the metallicity is primarily determined by the stellar mass, with a secondary dependence on the SFR. This secondary dependence on SFR is approximately 16\% times as strong as the primary dependence on stellar mass. As before, this dependence on the SFR is weaker than the dependence on the molecular gas content.

\begin{flushleft}
\textbf{Star formation efficiency}:
\end{flushleft}

The PCA results for SFE are shown in Fig. \ref{fig:vary_SFE}. $\sim 66\%$ of the variance is contained within the first component, and $\sim 97\%$ of the variance is contained within the first two. The first component is defined mainly in terms of stellar mass and SFE, with a slightly smaller contribution from metallicity. The second component is similarly defined, with the contribution from metallicity being a somewhat less. The third component, which here accounts for $3.0\%$ of the sample variation, is dominated by metallicity, with only a very minor contribution from SFE. 

Again, we set the third principle component to zero. Here, the third principle component is defined as 

\begin{multline}
{\rm PC 3} = 0.233 \;{\rm (M_*)^{PCA}} - 0.007 \;{\rm (SFE)^{PCA}} \\ -0.972 \; {\rm 12+log(O/H)^{PCA}}
\end{multline}

Setting this to zero implies that:

\begin{equation}
{\rm 12+log(O/H)} = 0.24 {\rm (log \; M_*)} - 0.008 {\rm (log \; SFE) + 6.45}
\end{equation}

That is, the metallicity is primarily determined by the stellar mass, with negligible secondary dependence on the SFE: the secondary dependence on SFE is approximately 3\% times as strong as the primary dependence on stellar mass. This clearly indicates that there is no correlation between SFE and metallicity, once the dependence on stellar mass is taken into account. This will be discussed further in the next section.

\begin{table}
\begin{tabular}{|l|c|c|c|}
\hline\hline
Component & M$_*$  & M(H$_2$) &  12+log(O/H) \\
\hline
PC 1   &  $      +0.578 \pm    0.040 $ &  $     -0.807 \pm    0.032 $  & $      +0.118 \pm    0.023 $\\
PC 2   &  $     -0.760 \pm     0.033$  &  $     -0.585 \pm    0.043 $ & $     -0.280 \pm     0.031 $  \\
PC 3   &  $      +0.295 \pm     0.027 $  &  $    -0.072 \pm     0.021$ & $     -0.952 \pm    0.008$ \\ \\
\hline
Component & M$_*$  & M(gas) &  12+log(O/H) \\
\hline
PC 1   &  $     +0.441\pm    0.081 $ &  $     -0.893\pm    0.041  $ &$     +0.085 \pm    0.023 $ \\
PC 2   &  $     -0.858 \pm    0.045 $ &  $     -0.448 \pm     0.083 $ & $     -0.249 \pm     0.022 $ \\
PC 3   &  $      +0.261 \pm     0.024 $ &  $    -0.037 \pm    0.015 $ & $     -0.964 \pm    0.007$ \\ \\
\hline
Component & M$_*$  & SFR  &  12+log(O/H) \\
\hline
PC 1   &   $+0.513 \pm 0.056$ &  $-0.852 \pm 0.036$ & $+0.100\pm0.027$\\
PC 3   &  $-0.816 \pm    0.036$ &  $-0.521 \pm 0.059$ & $-0.249 \pm0.021$ \\
PC 3   &  $+0.265 \pm     0.021 $ & $    -0.045 \pm     0.014$  &  $     -0.962 \pm    0.005$ \\ \\
\hline
Component & M$_*$  & SFE &  12+log(O/H) \\
\hline
PC 1    &  $      +0.904 \pm  0.042 $&  $     -0.368 \pm     0.114 $ & $      +0.214 \pm    0.023 $ \\
PC 2    &  $     -0.357 \pm     0.113$ &  $     -0.929 \pm    0.044 $ & $    -0.092 \pm    0.032 $ \\
PC 3    &  $      +0.233 \pm  0.025$ &  $   -0.007 \pm    0.004 $ & $     -0.972 \pm   0.006 $ \\
\hline\hline
\end{tabular}
\caption{Eigenvectors of the three principle components, for each of our parameter sets (stellar mass, metallicity, and a third parameter of interest -- respectively, molecular gas mass, total gas mass, SFR, and SFE. Errors have been obtained via a Monte Carlo bootstrapping method, as described in the text. }
\label{tab:PCA}
\end{table}

\section{Discussion}
\subsection{The effect of a changing CO/H$_2$ conversion factor}

Thus far, all principle component analysis results have been calculated using the metallicity-dependent CO/H$_2$ conversion factor presented by \cite{2010ApJ...716.1191W}. It is, however, important to investigate the extent to which these results are robust to a range of conversion factor prescriptions. 

We perform the same PCA steps for three further metallicity dependent CO/H$_2$ conversion factors: \cite{2011MNRAS.412..337G}, \cite{2012ApJ...747..124F}, and \cite{2012MNRAS.421.3127N}. In the interest of saving space, we do not repeat the above discussion for each case. Nevertheless, each conversion factor produces highly similar behaviour: examining each secondary-dependence in turn (molecular gas mass, {\it total} gas mass, SFR, and SFE), we find that the distribution of data does form a 2D plane in 3D parameter space, with between $2\% - 5\%$ of the sample variation being contained within the first two principle components. 

We focus on the differences produced by the  final step in the analysis -- that by setting the third principle component to zero, the metallicity can be defined  in terms of the stellar mass, with a secondary modifying term coming from one of the parameters listed above (in the original formulation of the FMR by \citealt{2010MNRAS.408.2115M} and \citealt{2010A&A...521L..53L}, SFR was used for this third parameter).

As above, we parameterise the strength of the effect of the third parameter (either M(H$_2$), M(gas), SFR, or SFE) on the metallicity by setting the third principle component to zero, and finding the resultant expression for the metallicity:

\begin{equation}
12+\log({\rm O/H}) = \log{\rm (M_*) - \mu  \log (X)},
\end{equation}

where X is, in turn, M(H$_2$), M(gas), SFR, or SFE. Fig. \ref{fig:vary_Xco} shows the values of $\mu$ we find for each parameter. All values of $\mu$ are $<1$, implying that in all cases the dominant driver of metallicity variations is the stellar mass.  For each of the four CO/H$_2$ conversion factors, we find that gas content is the most influential `secondary parameter'. For three out of four of the conversion factor prescriptions (\citealt{2010ApJ...716.1191W}, \citealt{2011MNRAS.412..337G}, and \citealt{2012MNRAS.421.3127N}), it is the molecular gas that has the most influence on metallicity (after stellar mass). For the remaining conversion factor prescription -- \cite{2012ApJ...747..124F} -- it is the {\it total} gas content that is the strongest secondary parameter (though it is comparable to the effect of molecular gas within the uncertainties on $\mu$, which are typically $\pm 0.05$). But, independent of the choice of conversion factor, we find that the strongest `FMR' effect is to be found with gas content.

\subsection{Comparison to the SFR-FMR}

The strength of SFR as a secondary parameter is somewhat less than gas content, with $\mu_{\rm SFR} \sim 0.17$ (small variations in $\mu_{\rm SFR}$ shown in Fig. \ref{fig:vary_Xco} are due to randomness introduced by the bootstrapping technique). We note that this value is significantly smaller than the value of $\mu$ found by \cite{2010MNRAS.408.2115M}: they express metallicity as a function of stellar mass and SFR:

\begin{equation}
12+\log({\rm O/H}) = \log{\rm (M_*) - 0.32 \log (SFR)}
\end{equation}

We therefore find the size of the effect of SFR on metallicity to be approximately 50\% of that found by \cite{2010MNRAS.408.2115M}. This discrepancy is far larger than can be accounted for by sample uncertainty; our bootstrapping analysis estimates a combined uncertainty on the influence of SFR of $\sim 30\%$ ($ \mu_{\rm SFR} = 0.17 \pm 0.05$).

One potential resolution to the conflict lies in the method used to calculate the SFRs. \cite{2010MNRAS.408.2115M} took their sample from SDSS, using the SFRs and strong line fluxes as provided by the SDSS spectroscopic fibre. SDSS calculates its pipeline SFRs by fitting to optical strong lines (with the H$\alpha$ having the most influence). As such (a) previously, SFRs and metallicities were calculated from matched-aperture observations, and (b) the same optical strong lines were used to calculate both SFRs and metallicities, which could introduce a co-dependency. Both of these effects could result in the correlation between SFR and metallicity being artificially tightened. It has been shown that when calculating SFRs by adopting {\it integrated} SFRs, corrected for fibre-aperture effects, the SFR-FMR effect weakens considerably (\citealt{2013MNRAS.433.1425B}; \citealt{2014ApJ...792...75Z}). 

In this work, we have used both galaxies with SDSS-SFRs (galaxies taken from the ALLSMOG and COLD GASS surveys), and galaxies with SFRs calculated with other methods (we used a combination of IR and UV fluxes for the HRS and LVL samples, the $z\sim2$ disks have SFRs derived from fitting to a combination of UV, optical continuum, H$\alpha$, and 24$\mu$m continuum fluxes, and the $z\sim2$ SMGs have SFRs derived using IR fluxes, via the far IR-radio correlation). These disparate methods, while all being effective probes of the total star formation rate, will clearly not introduce the same co-dependency as taking all values from the same SDSS spectra. This weaken any correlations between SFR and metallicity which were previously spuriously enhanced.


\subsection{Comparison with an analytical model}
\label{sec:SFE}

A key prediction of the model we discuss in \S3.2.1 is the existence of a `quenching sequence', populated by galaxies which are elevated {\it above} the mass-metallicity relation. Within the framework of the model, the only mechanism able to elevate galaxies significantly above the mass-metallicity relation is `strangulation' -- the cessation of gas infall -- which leads firstly to the build up of metals due to star formation (with a lack of metal-poor inflowing gas acting to dilute the ISM), and ultimately to the shutting down of star formation (`quenching'). Recently, Peng et al. (2015) have shown that the comparison of stellar metallicities between passive and star forming galaxies strongly support strangulation as the main quenching mechanism for galaxies with mass $<10^{11} {\rm M}_{\sun}$. Being an emission line-selected sample, none of our sample galaxies are currently fully quenched. The model predicts, however, that the $\sim 7$ galaxies elevated above our relation are essentially analogues of `green valley' galaxies, identified via their metallicity and gas mass, which have recently undergone strangulation and are in the processes of becoming quenched and dead.

It is important to investigate whether the model is doing a good job of explaining the scatter in the data. Randomness in the model is introduced by allowing the gas inflow rate to vary by a factor of two -- this introduces scatter in both metallicity and gas content. If it can be seen that the scatter in the observed data is greater than that contained in the model (allowing for observational uncertainties), it is likely that real gas inflow rates vary by a factor of $>2$. We compare the model to the data as follows. For each observed galaxy, we identify all model galaxies with the same stellar mass and metallicity (within bins of 0.05 dex). We then compare the total gas mass of the real galaxy to the mean gas mass of the (mass \& metallicity matched) model galaxies. We find a mean discrepancy between the model and observed gas mass of 0.49 dex. This must be compared, though, to the variance in gas mass contained within the model itself. Within each 0.05 dex bin of stellar mass and metallicity, the standard deviation of model gas masses is 0.41 dex. That is, the mean discrepancy between the analytical model and the data is only slightly more than the variance contained within the model itself. Allowing for typical observational uncertainties on the gas mass the model provides a good match to the data, suggesting that the assumptions of the model (i.e., that gas inflow rate varies by up to a factor of 2) are sufficient to explain the observed scatter.

\subsection{The connection between metallicity and star formation efficiency}
\label{sec:SFE}
\begin{figure}
\centering
\mbox
{
\includegraphics[clip=true, trim=0 0 0 0, width=8cm]{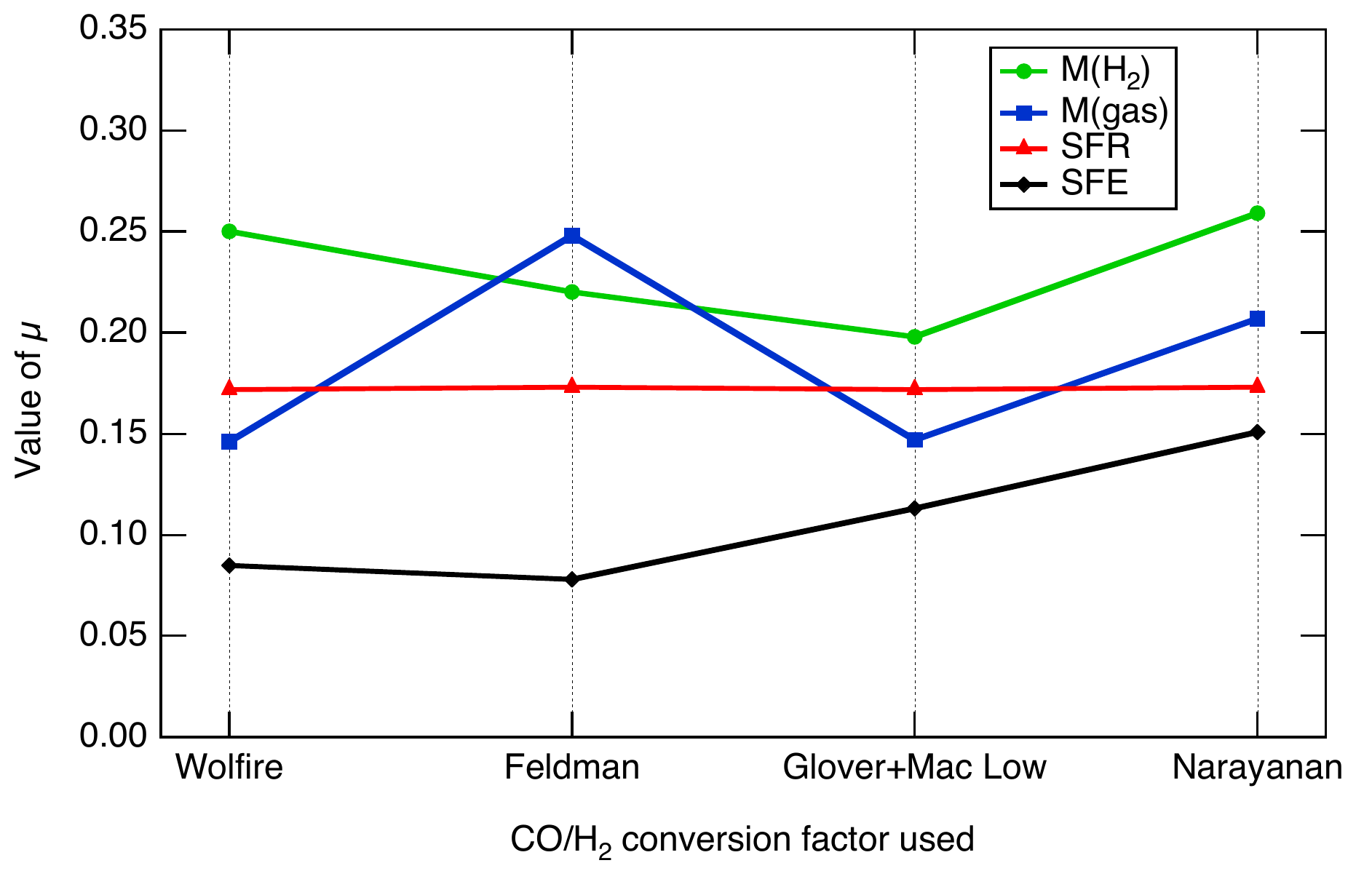}	
}
\caption{The strength of the influence on the mass-metallicity relation of molecular gas mass (green line), total gas mass (blue line), SFR (red line), and SFE (black line). The strength of the influence is quantified as $\mu$ -- see Eq.  12. Results for four recent metallicity-dependent CO/H$_2$ conversion factors are shown. In each case, the strongest effect is found to be with gas content (either molecular or total). The smallest influence on the mass-metallicity relation is found with star formation efficiency. 
}
\label{fig:vary_Xco}
\end{figure}

In all cases, by far the weakest `FMR' effect is found with the star formation efficiency. Across all CO/H$_2$ conversion factors, the SFE and metallicity are only very weakly connected. In other words, the Schmidt-Kennicutt relation does not depend on metallicity.  To date, the relationship between metallicity and star formation efficiency has been challenging to examine, due to the strong correlation between metallicity and stellar mass. The molecular gas consumption time ($\tau_{H2} \equiv 1/$SFE) is observed to vary with stellar mass (\citealt{2011MNRAS.415...61S}; \citealt{2014MNRAS.445.2599B}), a phenomenon which \cite{2011MNRAS.415...61S} attribute to the presence of increasingly `bursty' star formation histories towards lower halo masses. It is due to the strong co-variance between metallicity and stellar mass that a technique like Principle Component Analysis is required, in order to break the mass-metallicity degeneracy and isolate the effects of individual parameters. Simply plotting metallicity against star formation efficiency for our sample would result in a good, yet spurious, correlation.  

At first glance, a lack of correlation between metallicity and star formation efficiency appears somewhat surprising: metal lines act as an efficient cooling mechanism for the ISM, and given a lack of metals the ISM cannot effectively cool, fragment into molecular clouds, and form stars. This was the conclusion drawn by  \cite{2014Natur.514..335S}, who found extremely low star formation efficiencies in two local metal-poor dwarf galaxies (these low SFEs result from a large reservoir of CO-dark molecular gas, which \citealt{2014Natur.514..335S} trace using dust emission). However, the low metallicity regime in which the \cite{2014Natur.514..335S} study was carried out may differ significantly from the ISM conditions within the galaxies analysed in this work, and it is unclear if the results are applicable; many models predicting reduced SFE due to inefficient cooling in low metallicity environments  (i.e., \citealt{2009ApJ...699..850K}) deal with galaxies at such low metallicities that the ISM is entirely atomic. 


There are few explorations of the connection between SFE and metallicity in the molecular regime. \cite{2011MNRAS.415.3439D} present a semi-analytic model of protocluster clumps which predicts an {\it increase} in star formation efficiency as metallicity decreases, primarily due to the weaker stellar winds at low metallicity (which results in a less thorough evacuation of gas in low-metallicity star-forming clumps). Observationally, \cite{2011MNRAS.415.3439D} find a trend supporting this, which is, however, based on just three galaxies (the SMC, M33, and NGC6822). 

To date, therefore, there has been no clear observational consensus as to the effect of metallicity on star formation efficiency, with previous studies having very small sample sizes (2-3 galaxies) and reaching conflicting results.  
The results presented in this work, which find a negligible connection between star formation efficiency and metallicity, represent the first investigation of this issue with a significant sample size.

There are also many physical effects potentially governing the effect of metallicity on star formation efficiency (gas cooling and stellar wind efficiency being just two), and no clear theoretical consensus on which effects will dominate.  Phenomenologically, the weakness of the connection between SFE and metallicity follows simply from the fact that metallicity varies with molecular gas content and SFR in similar ways; a secondary dependence in the mass-metallicity relation is observed with both molecular gas and SFR (though molecular gas has the stronger effect). As this dependence is in the same direction for each parameter -- at a given stellar mass, galaxies with either increased SFR or increased M(H$_2$) have lower metallicity -- when considering the ratio of these two quantities (the SFE), the effect nearly cancels out. As a result, SFE and metallicity are only very weakly connected in the samples considered in this work. Physically, it may be that a combination of  metallicity-dependent processes (e.g. weak stellar winds, inefficient gas cooling) conspire to leave no connection between SFE and metallicity. Such a conclusion, however, would require theoretical investigation beyond the scope of this work.



\section{Conclusions}

In this work, we have presented an analysis of the connection between the gas content (both molecular and total) and metallicity of galaxies in samples from $0 < z < 2$. We have explored the effect of various metallicity-dependent CO/H$_2$ conversion factors on the Schmidt-Kennicutt relation. Conversely, we have also investigated the effect of gas content on the mass-metallicity relation, the most commonly used metallicity scaling relation. Finally, we use Principle Component Analysis to statistically analyse these parameters within a non-parametric framework. Our main conclusions are as follows:

\begin{itemize}


\item{We demonstrate the existence of a systematic `Fundamental Metallicity Relation' effect between molecular gas mass, stellar mass, and metallicity, by which at a fixed stellar mass, galaxies with higher H$_2$ mass lie systematically below the mass-metallicity relation. }\\

\item{We use Principle Component Analysis to explore the correlations between stellar mass, metallicity, and a third parameter (respectively, molecular gas content, total gas content, SFR, and SFE) in a non-parametric way. We find strong evidence for secondary dependences with both gas content and SFR, with the effect of gas content being stronger independent of the choice of CO/H$_2$ conversion factor. We conclude that gas content is the driver of the `Fundamental Metallicity Relation', with the (more commonly used) SFR-FMR being simply a consequence due to the Schmidt-Kennicutt relation. } \\

\item{We find a very negligible connection between gas-phase metallicity and star formation efficiency; that is, the Schmidt-Kennicutt relation has no dependence on metallicity. This result stands in opposition to some recent findings, which were conducted with far smaller sample sizes than used in this work. }

\end{itemize}

\section*{Acknowledgments}

We thank the anonymous referee who provided comments which improved this manuscript. This research has made use of NASA's Astrophysics Data System.

\bibliography{/Users/Matt/Documents/mybib}{}
\bibliographystyle{mn2e}

\appendix

\begin{table*}
\setcounter{table}{1}
\begin{tabular}{|c|c|c|c|c|}
\hline\hline
 log M$_*$            &  12+log(O/H) & SFR                           & M(H$_2$)             & Source\\
   {[M$_{\odot}$]} &                        &  [M$_{\odot}$/yr]  &  [M$_{\odot}$] &             \\
\hline
      9.70   &   8.89  &   1.70   &    8.27   &    ALLSMOG   \\
      9.61   &   9.15  &   2.32   &    9.20   &    ALLSMOG   \\
      9.58   &   8.98  &   1.64   &    7.85   &    ALLSMOG   \\
      9.94   &   9.08  &   3.84   &    8.80   &    ALLSMOG   \\
      9.84   &   8.99  &   1.24   &    8.90   &    ALLSMOG   \\
      9.73   &   8.81  &   1.16   &    8.93   &    ALLSMOG   \\
      9.80   &   9.08  &   1.29   &    8.55   &    ALLSMOG   \\
      9.85   &   9.11  &   1.18   &    8.98   &    ALLSMOG   \\
      9.91   &   8.95  &   1.54   &    8.91   &    ALLSMOG   \\
      9.70   &   9.01  &   1.06   &    8.57   &    ALLSMOG   \\
      9.44   &   8.71  &   2.17   &    8.91   &    ALLSMOG   \\
      9.89   &   8.97  &   1.72   &    8.75   &    ALLSMOG   \\
      9.99   &   9.08  &   1.05   &    8.64   &    ALLSMOG   \\
      9.64   &   8.89  &   1.13   &    8.54   &    ALLSMOG   \\
      9.98   &   9.07  &   1.43   &    8.92   &    ALLSMOG   \\
      9.80   &   8.98  &   1.05   &    8.34   &    ALLSMOG   \\
      9.32   &   8.91  &   2.14   &    7.60   &    ALLSMOG   \\
      9.91   &   9.09  &   1.39   &    9.06   &    ALLSMOG   \\
      9.96   &   9.12  &  0.160   &    8.47   &    ALLSMOG   \\
      9.68   &   9.01  &  0.358   &    8.32   &    ALLSMOG   \\
      9.98   &   9.03  &   1.89   &    8.52   &    ALLSMOG   \\
      9.74   &   8.92  &   1.79   &    8.52   &    ALLSMOG   \\
      9.57   &   8.80  &   3.90   &    8.36   &    ALLSMOG   \\
      9.78   &   9.00  &   1.81   &    8.68   &    ALLSMOG   \\
      9.36   &   8.91  &   1.62   &   $<$7.75   &    ALLSMOG   \\
      9.11   &   8.61  &   8.45   &   $<$7.94   &    ALLSMOG   \\
      9.42   &   8.74  &   1.05   &   $<$8.40   &    ALLSMOG   \\
      8.61   &   8.66  & 0.0551   &   $<$8.33   &    ALLSMOG   \\
      8.75   &   8.64  &  0.151   &   $<$8.44   &    ALLSMOG   \\
      8.57   &   8.61  &  0.129   &   $<$8.42   &    ALLSMOG   \\
      9.96   &   8.85  &   1.30   &   $<$8.38   &    ALLSMOG   \\
      9.51   &   8.84  &   1.01   &   $<$8.55   &    ALLSMOG   \\
      9.95   &   8.85  &   1.20   &   $<$8.49   &    ALLSMOG   \\
      8.67   &   8.52  & 0.0902   &   $<$8.45   &    ALLSMOG   \\
      9.51   &   8.80  &   3.70   &   $<$8.52   &    ALLSMOG   \\
      8.89   &   8.60  &  0.140   &   $<$8.34   &    ALLSMOG   \\
      8.99   &   8.60  &  0.305   &   $<$8.22   &    ALLSMOG   \\
      8.81   &   9.00  &  0.314   &   $<$8.08   &    ALLSMOG   \\
      8.96   &   8.65  &   1.50   &   $<$8.04   &    ALLSMOG   \\
      8.54   &   8.55  & 0.0357   &   $<$8.50   &    ALLSMOG   \\
      9.06   &   8.79  &   12.0   &   $<$7.95   &    ALLSMOG   \\
            10.17   &   8.95   &   2.11   &   8.87  &    COLD GASS   \\
      10.05   &   8.93   &   3.47   &   8.87       &    COLD GASS   \\
      10.08   &   8.99   &   2.19   &   8.91       &    COLD GASS   \\
      10.81   &   9.15   &   1.35   &   9.14       &    COLD GASS   \\
      10.74   &   9.17   &   4.41   &   9.13       &    COLD GASS   \\
      10.67   &   9.07   &   7.40   &   9.73       &    COLD GASS   \\
      10.74   &   9.07   &   9.33   &   9.85       &    COLD GASS   \\
      10.90   &   9.10   &   6.59   &   9.81       &    COLD GASS   \\
      10.26   &   9.08   &  0.910   &   8.36      &     COLD GASS   \\
      10.41   &   9.15   &   2.48   &   8.80       &    COLD GASS   \\
      10.55   &   9.13   &   2.41   &   9.14       &    COLD GASS   \\
      10.37   &   9.03   &   3.40   &   9.21       &    COLD GASS   \\
      10.48   &   9.10   &   1.13   &   9.03       &    COLD GASS   \\
      10.10   &   9.09   &   1.38   &   8.96       &    COLD GASS   \\
      10.53   &   9.13   &   26.8   &   9.30       &    COLD GASS   \\
      10.07   &   8.88   &   2.18   &   8.85       &    COLD GASS   \\
\hline\hline
\end{tabular}
\caption{Physical parameters (stellar mass, metallicity,  SFR, and M(H$_2$) derived using the Wolfire et al. (2010) CO/H$_2$ conversion factor) for the galaxies in this work.}
\label{tab:PCA}
\end{table*}  

\begin{table*}
\setcounter{table}{1}
\begin{tabular}{|c|c|c|c|c|}
\hline\hline
 log M$_*$            &  12+log(O/H) & SFR                           & M(H$_2$)             & Source\\
   {[M$_{\odot}$]} &                        &  [M$_{\odot}$/yr]  &  [M$_{\odot}$] &             \\
\hline
      10.11   &   9.08   &   2.42   &   9.15      &     COLD GASS   \\
      10.55   &   9.06   &   9.75   &   9.59      &     COLD GASS   \\
      10.37   &   9.04   &   8.57   &   9.46      &     COLD GASS   \\
      10.06   &   9.00   &   1.94   &   8.97      &     COLD GASS   \\
      10.11   &   9.08   &   2.53   &   8.93      &     COLD GASS   \\
      10.20   &   9.00   &  0.810   &   8.90     &      COLD GASS   \\
      10.38   &   9.10   &   2.89   &   9.35      &     COLD GASS   \\
      10.01   &   9.07   &   1.28   &   8.68      &     COLD GASS   \\
      10.32   &   8.87   &   10.0   &   9.58      &     COLD GASS   \\
      10.98   &   9.07   &   3.75   &   9.84      &     COLD GASS   \\
      10.27   &   9.02   &   1.00   &   9.17      &     COLD GASS   \\
      10.64   &   9.09   &   3.86   &   9.52      &     COLD GASS   \\
      10.03   &   8.95   &   1.24   &   9.49      &     COLD GASS   \\
      10.59   &   9.05   &   7.31   &   9.63      &     COLD GASS   \\
      10.24   &   9.10   &  0.110   &   8.75     &      COLD GASS   \\
      10.02   &   8.98   &  0.990   &   8.60     &      COLD GASS   \\
      10.39   &   9.04   &   3.38   &   9.29      &     COLD GASS   \\
      10.44   &   9.00   &   16.0   &   9.51      &     COLD GASS   \\
      10.80   &   9.17   &   2.72   &   9.18      &     COLD GASS   \\
      10.29   &   9.13   &   1.45   &   8.80      &     COLD GASS   \\
      10.98   &   9.15   &  0.620   &   8.91     &      COLD GASS   \\
      10.70   &   9.14   &   3.61   &   9.53      &     COLD GASS   \\
      10.09   &   9.05   &   5.72   &   9.23      &     COLD GASS   \\
      10.17   &   9.07   &   2.41   &   8.81      &     COLD GASS   \\
      10.77   &   9.07   &   5.51   &   9.68      &     COLD GASS   \\
      10.91   &   9.05   &   7.11   &   9.39      &     COLD GASS   \\
      10.60   &   9.18   &   3.46   &   9.07      &     COLD GASS   \\
      10.49   &   9.05   &   13.7   &   9.72      &     COLD GASS   \\
      10.15   &   9.03   &   2.11   &   8.86      &     COLD GASS   \\
      10.12   &   9.06   &   2.03   &   9.02      &     COLD GASS   \\
      10.77   &   9.04   &   6.84   &   9.76      &     COLD GASS   \\
      10.28   &   9.00   &   9.40   &   9.63      &     COLD GASS   \\
      11.03   &   9.08   &   4.80   &   9.61      &     COLD GASS   \\
      10.85   &   8.97   &   3.50   &   9.61      &     COLD GASS   \\
      10.95   &   9.19   &   3.47   &   9.63      &     COLD GASS   \\
      10.15   &   9.05   &   2.60   &   9.17      &     COLD GASS   \\
      10.99   &   8.92   &  0.410   &   9.97     &      COLD GASS   \\
      10.31   &   9.04   &   4.00   &   9.28      &     COLD GASS   \\
      10.18   &   8.96   &   9.73   &   9.42      &     COLD GASS   \\
      10.44   &   9.15   &   9.92   &   9.49      &     COLD GASS   \\
      11.33   &   9.14   &   1.92   &   9.35      &     COLD GASS   \\
      10.91   &   9.08   &   2.77   &   9.74      &     COLD GASS   \\
      10.41   &   9.15   &   3.22   &   9.10      &     COLD GASS   \\
      10.18   &   8.92   &   1.01   &   9.02      &     COLD GASS   \\
      10.76   &   9.07   &   5.74   &   9.56      &     COLD GASS   \\
      10.42   &   9.20   &   4.50   &   9.34      &     COLD GASS   \\
      10.77   &   9.05   &   4.72   &   9.73      &     COLD GASS   \\
      10.87   &   9.08   &   5.43   &   9.53      &     COLD GASS   \\
      10.57   &   9.10   &   1.93   &   8.96      &     COLD GASS   \\
      10.54   &   9.16   &   1.46   &   8.87      &     COLD GASS   \\
      11.03   &   9.18   &   1.47   &   8.91      &     COLD GASS   \\
      10.46   &   9.09   &  0.240   &   8.51     &      COLD GASS   \\
      10.28   &   9.13   &   2.41   &   8.77      &     COLD GASS   \\
      10.04   &   9.06   &   2.40   &   9.04      &     COLD GASS   \\
      10.02   &   8.99   &  0.350   &   8.43     &      COLD GASS   \\
      10.05   &   9.03   &   1.24   &   8.90      &     COLD GASS   \\
      10.56   &   9.16   &   1.72   &   8.67      &     COLD GASS   \\
      10.64   &   9.14   &   7.00   &   9.69      &     COLD GASS   \\
      10.10   &   8.98   &   2.14   &   9.27      &     COLD GASS   \\
      10.07   &   9.10   &   1.43   &   8.59      &     COLD GASS   \\
      10.13   &   9.06   &   1.57   &   8.98      &     COLD GASS   \\
      10.05   &   9.03   &   1.68   &   8.90      &     COLD GASS   \\
      10.60   &   9.02   &   3.44   &   9.51      &     COLD GASS   \\
      10.25   &   9.14   &   1.45   &   8.67      &     COLD GASS   \\
\hline\hline
\end{tabular}
\caption{(cont) Physical parameters (stellar mass, metallicity,  SFR, and M(H$_2$) derived using the Wolfire et al. (2010) CO/H$_2$ conversion factor) for the galaxies in this work. }
\label{tab:PCA}
\end{table*}

\begin{table*}
\setcounter{table}{1}
\begin{tabular}{|c|c|c|c|c|}
\hline\hline
 log M$_*$            &  12+log(O/H) & SFR                           & M(H$_2$)             & Source\\
   {[M$_{\odot}$]} &                        &  [M$_{\odot}$/yr]  &  [M$_{\odot}$] &             \\
\hline
      10.09   &   8.96   &   0.900  &   $<$8.88     &       COLD GASS   \\
      11.18   &   9.14   &    2.25  &   $<$9.17      &      COLD GASS   \\
      10.24   &   9.04   &   0.350  &   $<$8.93     &       COLD GASS   \\
      10.01   &   9.20   &    2.17  &   $<$9.10      &      COLD GASS   \\
      10.03   &   9.12   &  0.0600  &   $<$8.83    &        COLD GASS   \\
      10.09   &   8.94   &  0.000   &   $<$8.62     &       COLD GASS   \\
      11.01   &   9.12   &   0.520  &   $<$9.16     &       COLD GASS   \\
      10.89   &   9.17   &  0.0400  &   $<$9.09    &        COLD GASS   \\
      11.13   &   9.19   &   0.430  &   $<$9.25     &       COLD GASS   \\
      10.79   &   9.14   &    1.19  &   $<$9.12      &      COLD GASS   \\
      11.08   &   9.15   &   0.790  &   $<$9.18     &       COLD GASS   \\
      10.11   &   8.95   &  0.0600  &   $<$8.65    &        COLD GASS   \\

      9.46   &   8.81   &   3.04   &   8.89   &   Herschel Reference Survey   \\
      9.69   &   8.94   &   6.00   &   8.67   &   Herschel Reference Survey   \\
      10.2   &   9.02   &   13.8   &   9.34   &   Herschel Reference Survey   \\
      10.2   &   8.96   &   16.5   &   9.23   &   Herschel Reference Survey   \\
      9.77   &   8.76   &   8.34   &   9.16   &   Herschel Reference Survey   \\
      10.0   &   9.01   &   5.02   &   9.10   &   Herschel Reference Survey   \\
      9.42   &   8.83   &   3.44   &   8.45   &   Herschel Reference Survey   \\
      9.25   &   8.68   &   1.05   &   8.62   &   Herschel Reference Survey   \\
      9.28   &   8.85   &   1.80   &   8.47   &   Herschel Reference Survey   \\
      9.20   &   8.79   &   1.10   &   8.34   &   Herschel Reference Survey   \\
      9.21   &   8.62   &   2.42   &   8.39   &   Herschel Reference Survey   \\
      9.21   &   8.57   &   3.66   &   8.47   &   Herschel Reference Survey   \\
      9.22   &   8.69   &   2.86   &   8.78   &   Herschel Reference Survey   \\
      10.4   &   9.05   &   15.9   &   9.66   &   Herschel Reference Survey   \\
      9.47   &   8.68   &   2.19   &   8.43   &   Herschel Reference Survey   \\
      9.60   &   8.84   &   3.83   &   8.33   &   Herschel Reference Survey   \\

      9.45   &   8.84   &   1.55     &   $<$8.31   &   Herschel Reference Survey   \\
      9.17   &   8.74   &   7.54     &   $<$8.16   &   Herschel Reference Survey   \\
      8.94   &   8.92   &   2.04     &   $<$8.79   &   Herschel Reference Survey   \\
      9.27   &   8.59   &   3.63     &   $<$8.56   &   Herschel Reference Survey   \\
      9.20   &   8.82   &   0.701    &   $<$8.47   &   Herschel Reference Survey   \\
      9.23   &   8.73   &   1.61     &   $<$7.98   &   Herschel Reference Survey   \\
      9.14   &   8.60   &   0.983    &   $<$8.71   &   Herschel Reference Survey   \\
      
            11.48   &   8.99   &   4.29   &   9.36  &    LVL   \\
      9.736   &   8.36   &   0.37   &   7.85     &      LVL   \\
      10.86   &   8.90   &   2.45   &   9.22     &      LVL   \\
      10.83   &   9.31   &   5.20   &   9.12     &      LVL   \\
      9.485   &   8.64   &   0.13   &   7.22     &      LVL   \\
      8.407   &   8.01   &   0.05   &   8.14     &      LVL   \\
      8.450   &   8.01   &   0.07   &   8.19     &      LVL   \\
      9.976   &   8.76   &   0.37   &   8.35     &      LVL   \\
      10.74   &   9.09   &   0.97   &   8.67     &      LVL   \\
      11.12   &   9.04   &   0.65   &   8.66     &      LVL   \\
      11.10   &   9.13   &   3.10   &   9.51     &      LVL   \\
      8.582   &   8.23   &   0.04   &   7.76     &      LVL   \\
      9.722   &   8.65   &   0.17   &   6.98     &      LVL   \\
      8.769   &   8.25   &   0.16   &   7.04     &      LVL   \\
      9.296   &   8.32   &   0.31   &   8.78     &      LVL   \\
      9.611   &   8.80   &   0.11   &   7.21     &      LVL   \\
      11.01   &   8.89   &   1.91   &   8.77     &      LVL   \\
      9.382   &   8.31   &   0.74   &   7.92     &      LVL   \\
      9.376   &   8.64   &   0.28   &   7.76     &      LVL   \\
      10.64   &   9.01   &   0.57   &   8.59     &      LVL   \\
      10.45   &   9.01   &   0.47   &   8.60     &      LVL   \\
      
            10.47   &   9.05   &  245   &  10.75   &  MS \\
      10.43   &   8.95   &  30    &  10.44   &  MS \\
      9.780   &   8.73   &  33    &  10.85   &  MS \\
      10.75   &   8.73   &  97    &  10.70   &  MS \\
\hline\hline
\end{tabular}
\caption{(cont) Physical parameters (stellar mass, metallicity,  SFR, and M(H$_2$) derived using the Wolfire et al. (2010) CO/H$_2$ conversion factor) for the galaxies in this work.  }
\label{tab:PCA}
\end{table*}  

\begin{table*}
\setcounter{table}{1}
\begin{tabular}{|c|c|c|c|c|}
\hline\hline
 log M$_*$            &  12+log(O/H) & SFR                           & M(H$_2$)             & Source\\
   {[M$_{\odot}$]} &                        &  [M$_{\odot}$/yr]  &  [M$_{\odot}$] &             \\
\hline
      10.75   &   8.84   &  127   &  11.07   &  MS \\
      11.27   &   8.61   &  141   &  11.07   &  MS \\
      11.38   &   8.96   &  117   &  11.02   &  MS \\
      11.17   &   8.98   &  92    &  10.80   &  MS \\
      11.23   &   9.01   &  141   &  11.17   &  MS \\

      10.79  &    8.75  &    540   &    11.54 &   SMG \\
      11.20  &    9.09  &    810   &    11.36 &   SMG \\   
      10.30  &    8.75  &    1070  &    11.02 &   SMG \\
      10.86  &    8.74  &    390   &    11.48 &   SMG \\
      10.59  &    8.91  &    680   &    10.72 &   SMG \\
      10.99  &    9.07  &    2890  &    11.49 &   SMG \\
      11.01  &    9.09  &    1260  &    11.39 &   SMG \\
      11.13  &    8.86  &    950   &    11.07 &   SMG \\
\hline\hline
\end{tabular}
\caption{(cont) Physical parameters (stellar mass, metallicity,  SFR, and M(H$_2$) derived using the Wolfire et al. (2010) CO/H$_2$ conversion factor) for the galaxies in this work.  }
\label{tab:PCA}
\end{table*}  

\end{document}